\documentclass[fleqn,usenatbib]{mnras}

\usepackage{newtxtext,newtxmath}
\usepackage{graphicx}	
\usepackage{amsmath}

\usepackage[T1]{fontenc}
\usepackage{caption}
\usepackage{subcaption}
\definecolor{peru}{RGB}{205,133,63}
\definecolor{dodgerblue}{RGB}{30,144,255}
\definecolor{royalblue}{RGB}{65,105,225}
\definecolor{deeppink}{RGB}{255,20,147}
\definecolor{darkviolet}{RGB}{148,0,211}
\definecolor{slategray}{RGB}{112,128,144}
\definecolor{yellowgreen}{RGB}{154,205,50}
\definecolor{crimson}{RGB}{220,20,60}
\definecolor{orangered}{RGB}{255,69,0}
\definecolor{sienna}{RGB}{160,82,45}
\definecolor{dodgerblue}{RGB}{30,144,255}
\definecolor{darkviolet}{RGB}{148,0,211}
\definecolor{slategray}{RGB}{112,128,144}
\definecolor{hotpink}{RGB}{255,105,180}
\definecolor{yellowgreen}{RGB}{154,205,50}
\definecolor{olive}{RGB}{128,128,0}
\definecolor{crimson}{RGB}{220,20,60}
\definecolor{deeppink}{RGB}{255,20,147}
\definecolor{blueviolet}{RGB}{138,43,226}
\definecolor{limegreen}{RGB}{50,205,50}
\definecolor{springgreen}{RGB}{0,255,127}
\definecolor{teal}{RGB}{0,128,128}
\definecolor{peru}{RGB}{205,133,63}
\definecolor{deepskyblue}{RGB}{0,191,255}
\definecolor{royalblue}{RGB}{65,105,225}
\definecolor{gold}{RGB}{255,215,0}
\definecolor{cyan}{RGB}{0,255,255}
\definecolor{magenta}{RGB}{255,0,255}
\usepackage[table]{xcolor}

\DeclareRobustCommand{\VAN}[3]{#2}
\let\VANthebibliography\thebibliography
\def\thebibliography{\DeclareRobustCommand{\VAN}[3]{##3}\VANthebibliography}

\usepackage{graphicx}	
\usepackage{amsmath}	
\usepackage{orcidlink}
\usepackage{multirow}

\title[The $\bar{\mathcal{I}}-\mathcal{C}$ relations in $f(Q)$ gravity]{Exploring the universal $\bar{\mathcal{I}}-\mathcal{C}$ relations for relativistic stars in $f(Q)$ gravity}

\author[Alwan et al.]{
Muhammad Azzam Alwan $^{\orcidlink{0000-0002-0558-2092}}$,$^{1,2}$\thanks{E-mail:azzam-alwan@hiroshima-u.ac.jp}
Tomohiro Inagaki $^{\orcidlink{0000-0003-2777-4017}}$,$^{1,3,4}$\thanks{E-mail:inagaki@hiroshima-u.ac.jp}
S.A. Narawade $^{\orcidlink{0000-0002-8739-7412}}$,$^{5}$\thanks{E-mail:shubhamn2616@gmail.com}
B. Mishra $^{\orcidlink{0000-0001-5527-3565}}$,$^{5}$\thanks{E-mail:bivu@hyderabad.bits-pilani.ac.in}
\\
$^{1}$Graduate School of Advanced Science and Engineering, Hiroshima University, Higashi-Hiroshima 739-8526, Japan.\\
$^{2}$Research Center for Quantum Physics, National Research and Innovation Agency (BRIN), Tangerang Selatan 15314, Indonesia.\\
$^{3}$Information Media Center, Hiroshima University, Higashi-Hiroshima 739-8521, Japan.\\
$^{4}$Core of Research for the Energetic Universe, Hiroshima University, Higashi-Hiroshima 739-8526, Japan.\\
$^{5}$Department of Mathematics, Birla Institute of Technology and Science-Pilani,Hyderabad Campus, Hyderabad-500078, India.
}

\date{Accepted XXX. Received YYY; in original form ZZZ}

\pubyear{\the\year{}}

\begin{document}
\label{firstpage}
\pagerange{\pageref{firstpage}--\pageref{lastpage}}
\maketitle

\begin{abstract}
We investigate the properties of neutron stars within the framework of $f(Q)$ gravity by incorporating rotational effects through a slowly rotating metric. We derive the modified TOV equations and calculate the angular velocity profiles and moments of inertia (MOI) for linear, quadratic, exponential, and logarithmic $f(Q)$ models. Our results show that deviations in the MOI are more pronounced than those in the stellar mass profiles, suggesting that rotational observables are highly sensitive to geometric corrections. We also calculate a quasi-universal relation between the dimensionless MOI and compactness ($\bar{I}$-$C$). The linear and quadratic models are generally consistent with observational data from PSR J0737-3039A, although the deviations are small and difficult to distinguish from General Relativity due to inherent EoS variability. On other hand, the logarithmic and exponential models show larger deviations (over 20\%), exceeding the EoS-induced uncertainty reported by Suleiman \&
Read (2024), highlighting the relation's sensitivity to the $f(Q)$ gravity model. These results indicate that $f(Q)$ gravity could potentially be tested in the strong-field regime and point to a direction for future studies, such as investigating EoS-insensitive quasi-universal relations, like the $\bar{I}(\Lambda)$ relations, within the $f(Q)$ framework. Such relations may provide a clearer pathway for exploring possible signatures in strong-field gravity when combined with more precise future observations.
\end{abstract}




\section{Introduction}\label{Sec1}
General Relativity (GR) has been proven to be a successful theory (\cite{Baker:2014zba}), extensively tested across wide range of environments. These include weak-field scenarios such as the Solar System and laboratory experiments (\cite{Bambi:2024kqz}), as well as strong-field regimes. For example, binary pulsars such as the double pulsar system PSR J0737–3039 A/B have provided tests of GR in the strong-field regime, confirming predictions such as gravitational wave damping, Shapiro delay, and periastron advance with high precision (\cite{Kramer:2021jcw}). Another avenue of strong-field tests comes from the prediction and observation of Gravitational Waves (GW) (\cite{LIGOScientific:2016lio, LIGOScientific:2018dkp, LIGOScientific:2020tif}). The GW detected by LIGO and Virgo are waves produced by the interaction of two compact objects in astrophysics, such as black holes (BHs) and neutron stars (NS), which are high-density objects with extreme gravitational environments. The first GW detection, GW150914 (\cite{LIGOScientific:2016vbw}), was produced by a black hole merger. This event not only provided strong support for GR, but also opened up the opportunity to test gravity under extreme conditions, where deviations from GR can occur, allowing for the testing of alternative theories of gravity. However, testing gravity through black holes has limitations, mainly because of the difficulty in obtaining observational data. Unlike black holes, neutron stars have abundant observational data and have interior structures rich in matter such as degenerate neutrons or various exotic materials, such as hyperons, quarks, or pion condensates, making them ideal astrophysical objects to test gravity in strong fields (\cite{Shao:2022koz}).

One approach to calculate the properties of a neutron star is by solving the Tolman-Oppenheimer-Volkoff (TOV) equation, incorporating a given Equation of State (EoS), and then comparing the theoretical predictions with various observational constraints. Among these constraints are the mass-radius measurements obtained from pulsar observations, particularly those provided by the Neutron Star Interior Composition Explorer (NICER).
For example, the millisecond pulsar PSR J0030+0451 has been analyzed independently by two groups: \cite{Miller2019} reported a mass of $1.34^{+0.15}_{-0.16} \, M_\odot$ with a radius of $12.71^{+1.14}_{-1.19}$ km, while \cite{Riley2019} found a mass of $1.44^{+0.15}_{-0.14} \, M_\odot$ and a radius of $13.02^{+1.24}_{-1.06}$ km. These differences arise from distinct modeling assumptions concerning the thermal X-ray emission from the neutron star’s surface hot spots. Another massive pulsar observed by NICER, PSR J0740+6620, provides further constraints. Using NICER and X-ray Multi-Mirror(XMM)-Newton data, \cite{Miller2021} obtained a mass of $2.08 \pm 0.07 \, M_\odot$ and a radius of $13.7^{+2.6}_{-1.5}$ km, while \cite{Riley2021} derived $2.072^{+0.067}_{-0.066} \, M_\odot$ and $12.39^{+1.30}_{-0.98}$ km, employing informative priors from NANOGrav and CHIME/Pulsar timing measurements.

In addition to these X-ray measurements, radio pulsar timing provides the most precise and least model-dependent mass constraints. The detection of a strong Shapiro delay in PSR J1614-2230 led to a mass measurement of $1.97 \pm 0.04 \, M_\odot$~(\cite{Demorest:2010bx}), providing the first robust evidence for a nearly two-solar-mass neutron star. This discovery ruled out a broad class of soft EoS models that predict strong softening due to the presence of hyperons or meson condensates at supranuclear densities, while allowing for quark matter only under the assumption of strong interactions. Some years later, the refined timing of PSR J0740+6620 yielded a mass of $2.08 \pm 0.07 \, M_\odot$~(\cite{Fonseca:2021wxt}), currently regarded as the most reliable lower limit on the maximum neutron star mass. Together, these results require that any viable EoS must be sufficiently stiff to support neutron stars with masses of at least $2\,M_\odot$, establishing an essential benchmark for testing dense matter physics and strong-field gravity. Other massive pulsars also provide important insight into the stiffness of the EoS. For example, PSR J2215+5135, one of the most massive neutron stars known, has a measured mass of $2.27^{+0.17}_{-0.15} \, M_\odot$, obtained through combined radio and optical spectroscopy observations~(\cite{Linares:2018ppq}). Such detections have already ruled out a range of soft EoS models that cannot align their predictions with the observed NS mass. In addition, the double pulsar system PSR J0737--3039A/B~(\cite{Burgay:2003jj,Lyne:2004cj}) serves as a benchmark for both gravity tests and dense matter studies. Long-term timing of this system has allowed the measurement of all five post-Keplerian parameters with exquisite precision~(\cite{Kramer:2021jcw}), yielding $M_A = 1.3381\,M_\odot$ and $M_B = 1.2489\,M_\odot$. This pulsar provides another constraint on the EoS through the calculation of the moment of inertia~(\cite{Lattimer:2004nj,Worley:2008cb,Landry:2018jyg,Lim:2018xne,Jiang:2020uvb,Miao_2021_515_5071}). The same dataset is expected to enable a direct determination of the moment of inertia of PSR J0737-3039A within the next decade, with an accuracy of order 10\%, offering a new probe of the neutron star EoS and an independent test of strong-field gravity.

However, using modified gravity theories, the predictions of GR deviate, offering an alternative where previously ruled-out EoS can match observed properties, such as higher or lower masses for stiff EoS (\cite{Yazadjiev:2014cza, Capozziello:2015yza, Babichev:2016jom, Astashenok:2021peo, Lin:2021ijx, Liu:2024wvw, Cui:2024nkr}). Amid this EoS uncertainty, Yagi and Yunes (2013) introduced several universal relations, namely the relations between different dimensionless quantities that are weakly dependent on the EoS (\cite{Yagi:2013bca}). This relation reveals a common pattern, even though the interaction between gravity and matter may vary in each EoS, thus providing a new perspective on testing theories in the strong-field regime. Various modifications of gravity have been studied to examine how they deviate from different universal relations in neutron stars, rather than due to changes in the EoS (\cite{Staykov2016, Sakstein_2017_95_064013, Boumaza:2021fns, Blazquez-Salcedo:2022pwc, Xu:2021kfh, Murshid:2023xsw}). This, of course, would serve as a valuable tool for testing both GR and alternative theories of gravity. In this paper, we will explore an universal relation for neutron stars within the $f(Q)$ gravity framework.

One of the main drivers of modified gravity research is the cosmological constant problem and the elusive nature of dark energy (\cite{Copeland_2006_15_1753, Clifton_2012_513_1, Joyce_2015_568_1, Koyama_2016_79_046902, Bull_2016_12_56}). The $f(Q)$ theories represent a natural extension of Einstein's GR and have gained significant attention as alternative gravity theories, particularly for explaining the accelerated expansion of the universe. These theories involve modifying the standard Einstein-Hilbert Lagrangian, replacing it with a function of the Non-metricity scalar $Q$. Various classes of $f(Q)$ theories have been explored, and many models have been constructed and analyzed (for a detailed review, see \cite{Jimenez_2018_98_044048, Harko_2018_98_084043, Anagnostopoulos_2021_822_136634, Anagnostopoulos_2023_83_58, Heisenberg_2023_83_315, Nojiri:2024zab,Heisenberg:2023lru}). While $f(Q)$ theories are primarily applied to cosmological phenomena (\cite{Khyllep_2021_103_103521, Koussour_2022_36_101051, Frusciante_2021_103_044021, Narawade_2022_36_101020, Narawade_2023_535_2200626, Heisenberg:2023tho}), they also have important implications for astrophysical objects. In the context of $f(Q)$ gravity, compact objects have been studied extensively in the literature~(\cite{DAmbrosio:2021zpm, Maurya_2022_70_2200061, Alwan_2024_2024_011, Bhar_2023_42_101322, Bhar_2024_43_101391, Lohakare_2023_526_3796, Gul_2024_84_8}). In our previous study~(\cite{Alwan_2024_2024_011}), we examined how the interior geometry of neutron stars is influenced by non-metricity under static condition, which subsequently affects their density, pressure, and stability. Deviations arising from modifications of $f(Q)$ can enable stars to accommodate more matter, increasing their mass, or lead to a loss of stability, causing them to become lighter or even collapse. In one of the $f(Q)$ models, we observed maximum mass deviations that could accommodate EoS like APR4, aligning with various observational constraints from pulsar and gravitational wave observations. 

Building on these results, we extended our study to neutron stars in $f(Q)$ gravity in the slowly rotating case, employing 53 different realistic EoS. The effects of slow rotation and the moment of inertia of anisotropic stars in $f(Q)$ gravity have been explored by \cite{Errehymy:2024qpu}. They used an empirical formula for the moment of inertia in terms of the total mass and radius for static solutions, as provided by \cite{Bejger:2002ty}. In this paper, we directly incorporate the effects of rotation into the metric, allowing us to derive moment of inertia properties that cannot be obtained in the static case.  From various references on other modified gravity theories, we expect that $f(Q)$ models can cause deviations in the angular velocity ($\bar{\omega}_c$) at the center of neutron stars, resulting in greater deviations in the moment of inertia compared to our previous calculations in the mass-radius diagram. Through the moment of inertia, we also calculate one of the universal relations between the dimensionless moment of inertia ($\bar{I}$) and the compactness of stars ($C$). Based on the differences in $\bar{\omega}_c$ values, we anticipate obtaining a $\bar{I}-C$ relation with a shape similar to that of GR, but shows different deviation between each model which shows their respective signatures. From these calculations, we expect this relationship to serve as a constraint on the $f(Q)$ parameter through possible future measurements.

This paper is organized as follows: In Section \ref{Sec2}, we introduce the mathematical framework of $f(Q)$ gravity, along with the numerical setup, which includes the step to solve the field equations and the selection of the EoS. Section \ref{Sec3} focuses on analyzing the impact of $f(Q)$ models on the properties of slowly rotating neutron stars, particularly focusing on the moment of inertia. In this section, we also investigate the universal $\bar{I}-C$ relation, computing and fitting it across various $f(Q)$ models. Finally, we summarize our findings and outline potential future directions in Section \ref{Sec4}.

\section{Mathematical Framework}\label{Sec2}
\subsection{Basic $f(Q)$ Formalism and Covariant Formulation}\label{Sec2A}
In a spacetime characterized by a metric tensor $g_{\mu\nu}$ and an affine connection $\Gamma_{~~\mu \nu}^\lambda$, the torsion tensor $T^{\lambda}_{~~\mu\nu}$ and the non-metricity tensor $Q_{\lambda\mu\nu}$ are defined as,
\begin{eqnarray}\label{eq1}
    T^{\lambda}_{~~\mu\nu} &:=& \Gamma^{\lambda}_{~~\mu\nu}-\Gamma^{\lambda}_{~~\nu\mu}~,\nonumber\\[5pt]
    Q_{\lambda\mu\nu} &:=& \nabla_{\lambda}g_{\mu\nu} = \partial_{\lambda}g_{\mu\nu}-\Gamma^{\alpha}_{~~\lambda\mu}g_{\alpha\nu}-\Gamma^{\alpha}_{~~\lambda\nu}g_{\alpha\mu}~.
\end{eqnarray}
The decomposition of the general affine connection allows us to clearly distinguish the contributions from torsion and non-metricity, separating them from the purely metric-compatible component of the connection. This can be represented as the sum of the Levi-Civita connection $\left(\left\{_{~~\mu \nu}^\lambda\right\}\right)$, contortion $\left(K_{~~\mu \nu}^\lambda\right)$, and disformation $\left( L_{~~\mu v}^\lambda\right)$. This relationship can be expressed mathematically as,
    \begin{equation}\label{eq2}
    \Gamma_{~~\mu \nu}^\lambda= \left\{_{~~\mu \nu}^\lambda \right\}+K_{~~\mu \nu}^\lambda+L_{~~\mu \nu}^\lambda~,
    \end{equation}
where
\begin{eqnarray*}
    \left\{_{~~\mu \nu}^\lambda \right\} &=& \frac{1}{2} g^{\lambda \alpha}\left(\partial_\mu g_{\alpha \nu}+\partial_\nu g_{\alpha \mu}-\partial_\alpha   g_{\mu \nu}\right),\nonumber\\[5pt]
    K_{~~\mu \nu}^\lambda &=& \frac{1}{2}\left(T_{~~\mu\nu}^\lambda+T_{\mu~~\nu}^{~~\lambda}+T_{\nu~~\mu}^{~~\lambda}\right),\nonumber\\[5pt]
    L_{\ \mu \nu }^{\lambda} &=& \frac{1}{2}(Q_{~~\mu\nu }^{\lambda}-Q_{\mu~~\nu }^{\ \lambda}-Q_{\nu~~\mu }^{\ \lambda})~.
\end{eqnarray*}
A modified $f(Q)$ theory of gravity with symmetric teleparallelism considers an affine connection with vanishing curvature and null torsion and let non-metricity be the only driving force. The non-metricity scalar is defined as \cite{Jimenez_2018_98_044048},
\begin{equation}\label{eq3}
    Q = Q_{\lambda\mu\nu}P^{\lambda\mu\nu} = \frac{1}{4}\Big( -Q_{\lambda\mu\nu}Q^{\lambda\mu\nu}+2Q_{\lambda\mu\nu}Q^{\mu\nu\lambda}+Q_{\lambda}Q^{\lambda}-2Q_{\lambda}\tilde{Q}^{\lambda}\Big)~.
\end{equation}
where $Q_{\lambda} = g^{\mu\nu}Q_{\lambda\mu\nu}$, $\tilde{Q}^{\lambda} = g^{\mu\nu}Q_{\mu\nu\lambda}$ are two traces of non-metricity tensor and its conjugate $P^{\lambda}_{~~\mu\nu}$ is called as superpotential, given by
\begin{align}\label{eq4}
    P^{\lambda}_{~~\mu\nu} = -\frac{1}{4}Q^{\lambda}_{~\mu \nu} +& \frac{1}{4}\Big(Q^{~\lambda}_{\mu~\nu} + Q^{~\lambda}_{\nu~~\mu}\Big) + \frac{1}{4}Q^{\lambda}g_{\mu \nu}\nonumber\\
    &-\frac{1}{8}\Big(2 \tilde{Q}^{\lambda}g_{\mu \nu} + {\delta^{\lambda}_{\mu}Q_{\nu} + \delta^{\lambda}_{\nu}Q_{\mu}} \Big)~.
\end{align}
The non-metricity scalar $Q$ is invariant under local general linear transformations and translational symmetries. An alternative definition found in the literature, $Q = -Q_{\lambda\mu\nu}P^{\lambda\mu\nu}$, results in a sign change for the scalar. This is important when comparing different $f(Q)$ results. Additionally, the non-metricity scalar $Q$ can replace the Ricci scalar $R$ in the Einstein-Hilbert action, yielding the symmetric teleparallel equivalent of General Relativity (GR), known as STEGR. It is noteworthy that symmetric teleparallel theory faces the same `dark' problem as conventional GR. Modified gravity theories of the form $f(Q)$ were developed to address these issues, in a similar way to the extensions seen in modified gravity theories represented by $f(R)$.

The components of the connection in Eq. \eqref{eq2} allow one to choose a coordinate system $\{y^\mu\}$ in which the affine connection $\Gamma^\lambda_{~\mu\nu}(y^\mu)$ vanishes, which is called the coincident gauge. Then in any other coordinate system $\{x^\mu\}$, the affine connection then takes the following form (\cite{Zhao_2022_82_303}).
\begin{equation}
\Gamma^\lambda_{~\mu\nu}(x^\mu) = \frac{\partial x^\lambda}{\partial y^\beta} ~ \partial_\mu \partial_\nu y^\beta.
\end{equation}
Since there exists a coordinate system $\{y^\mu\}$ in which the affine connection vanishes, one can always assume that calculations are being performed in this special coordinate system, where the metric is the only fundamental variable. Typically, fixing a specific coordinate system implies a breaking of diffeomorphism symmetry. However, this is not the case in the STEGR framework. At the outset, it is not known in which coordinate system the affine connection vanishes. Therefore, one may assume that in an arbitrary coordinate system $A$, the affine connection is zero. Alternatively, one could consider another coordinate system $B$ and similarly assume the affine connection vanishes there. It can be shown that the difference in the action between these two cases amounts only to a total derivative. Since surface terms do not affect the equations of motion, it follows that the evolution of the metric remains unchanged regardless of the coordinate system in which the affine connection is set to zero. This demonstrates that the theory preserves diffeomorphism symmetry despite the choice of coordinate system. In the coincident gauge, the covariant derivative $\nabla _{\lambda}$ simplifies to the partial derivative $\partial _{\lambda}$ leading to the expression $Q_{\lambda \mu \nu }=\partial _{\lambda}g_{\mu \nu }$. Therefore, the Levi-Civita connection $\left(\left\{_{~~\mu \nu}^\lambda\right\}\right)$ can be rewritten in terms of the disformation tensor $L_{\ \mu \nu }^{\alpha }$ as$\left(\left\{_{~~\mu \nu}^\lambda\right\}\right)=-L_{\ \mu \nu }^{\lambda}$. By varying the action term (\cite{Jimenez_2018_98_044048, Zhao_2022_82_303}),

\begin{equation}\label{eq6}
    S = \int \frac{1}{2\kappa}f(Q)\sqrt{-g}~d^{4}x + \int \mathcal{L}_{m}\sqrt{-g}~d^{4}x~,
\end{equation}
where, $f(Q)$ is an arbitrary function of nonmetricity $Q$, $\mathcal{L}_{m}$ be the matter Lagrangian density and $g$ is the determinant of the metric tensor $g_{\mu\nu}$. Now varying the action with respect to the metric tensor, we can obtain the field equation
\begin{align}\label{eq7}
  \frac{2}{\sqrt{-g}}&\nabla_{\lambda}\left(\sqrt{-g}f_{Q}P^{\lambda}_{~~\mu\nu}\right) - \frac{1}{2}g_{\mu \nu}f \nonumber\\
  &+ f_{Q}(P_{\mu\lambda\alpha}Q^{~~\lambda \alpha}_{\nu} - 2Q_{\lambda \alpha \mu}P^{\lambda \alpha}_{~~~\nu}) = \kappa \mathcal{T}_{\mu \nu}~.   
\end{align}
It is well established that the original STEGR theory yields the same dynamics as GR, and that the evolution of the metric remains unchanged regardless of the choice of affine connection (\cite{Zhao_2022_82_303}). However, the situation differs in modified versions of STEGR, such as $f(Q)$ gravity, where the nonmetricity scalar $Q$ in the STEGR action is replaced by a general function $f(Q)$. In such cases, the part of the action that depends on the affine connection is no longer a total derivative. As a result, different choices of affine connection can lead to different evolutions of the metric such as non-zero affine connection ($\Gamma^\lambda_{~\mu\nu}\neq 0$). Using this field equation, the covariant formulation has been developed and used effectively in studying geodesic deviations and cosmological phenomena (\cite{Lin_2021_103_124001, Zhao_2022_82_303, Beh_2022_77_1551, Subramaniam_2023_71_2300038}),

\begin{equation}\label{eq8}
    f_{Q}\mathring{G}_{\mu\nu}+\frac{1}{2}g_{\mu\nu}(Qf_{Q}-f)+2f_{QQ}P^{\lambda}_{~~\mu\nu}(\mathring\nabla_{\lambda} Q) = \kappa \mathcal{T}_{\mu \nu}~,
\end{equation}
where, $f_{Q}$ is derivative of $f$ with respect to $Q$, $\mathring\nabla$ is the covariant derivative associated with Levi-Civita connection and $\mathring{G}_{\mu\nu} = R_{\mu\nu}-\frac{1}{2}g_{\mu\nu}R$, with $R_{\mu\nu}$ and $R$ are the Riemannian Ricci tensor and scalar respectively which are constructed by the Levi-Civita connection. In the case of STEGR, where $f(Q) = Q$, the left hand side of Eq. \eqref{eq8} corresponds to the Einstein tensor, which depends solely on the metric. As discussed, that the choice of affine connection has no impact on the evolution of the metric. However, in more general scenarios where $f(Q)$ is not a linear function of $Q$, the affine connection explicitly enters the dynamical equations and therefore influences the evolution of the metric. Variation of Eq. \eqref{eq6} with respect to the connection, we can derive the equation of motion for the nonmetricity scalar as,
\begin{equation}\label{eq9}
\nabla_{\mu}\nabla_{\nu}\left(\sqrt{-}f_{Q}P^{\mu\nu}_{~~~\lambda}\right)=0~.
\end{equation}

\subsection{Numerical Setup}\label{Sec2B} 
We have developed a modified TOV equation for non-rotating neutron stars in covariant $f(Q)$ gravity across several models, providing profiles and properties in our previous work (\cite{Alwan_2024_2024_011}). Now, we will calculate slowly rotating neutron stars by incorporating rotational terms as (\cite{Hartle_1967_150_1005, Sakstein_2017_95_064013, Breu_2016_459_646})
\begin{equation}\label{eq10}
    ds^{2} = -e^{A(r)}dt^{2}+e^{B(r)}dr^{2}+r^{2}d\theta^{2}+r^{2}sin^{2}\theta d\phi^{2}-2\omega(r)\epsilon~r^{2}sin^{2}dt~d\phi~,
\end{equation}
This metric describes the spacetime geometry with a rotational term, $\omega(r)$, where $\omega(r)$ represents the angular velocity of the inertial frame dragged by the rotation of star. The parameter $\epsilon$ in the $\omega$ term denotes slow rotation, where $\epsilon$ must be small, $\epsilon \ll 1$. When $\epsilon$ equals zero, we recover the static and spherically symmetric metric. Now, by eliminating the parameters that contain information of gravity in the metric Eq. \eqref{eq10}, we can achieve a flat, curvature-free spacetime by simply setting $A(r) = 0$, $B(r) = 0$, and $\epsilon = 0$. This reduces the metric to the Minkowski metric in spherical coordinates:
 \begin{equation}\label{eq11}
    ds^{2} = -dt^{2}+dr^{2}+r^{2}d\theta^{2}+r^{2}sin^{2}\theta d\phi^{2}~,
\end{equation}
and denoted by $\mathcal{G}_{(r)}$. Our next step is to determine which affine connection best fits the spacetime determined by the metric $\mathcal{G}_{(r)}$. Further, it is well known in GR that gravity in Minkowski spacetime Eq. \eqref{eq11} is represented by the curvature tensor $\mathring{R}_{(r)\mu\nu\rho}^{~~~~~~~~~\sigma}$ and in STEGR theory, it is represented by the non-metricity tensor $Q_{(r)\lambda\mu\nu}$~, so it is meaningful to assume $Q_{(r)\lambda\mu\nu}=0$ for the new spacetime determined by $\mathcal{G}_{(r)}$. Based on this assumption and Levi-Civita connection Eq. \eqref{eq2}, we can calculate the non-vanishing components of arbitrary affine connections:
\begin{eqnarray}\label{eq12}
 \Gamma^\theta_{~r\theta} &=& \Gamma^\theta_{~\theta r} = \Gamma^\phi_{~r\phi} = \Gamma^\phi_{~\phi r} = \frac{1}{r}, \quad \Gamma^{\theta}_{~\phi\phi} = -\cos\theta\sin\theta, \nonumber\\
 \Gamma^\phi_{~\theta\phi} &=& \Gamma^\phi_{~\phi\theta} = \cot\theta, ~~~ \Gamma^r_{~\phi\phi} = -r\sin^2\theta, ~~~ \Gamma^{r}_{~\theta\theta} = -r.
\end{eqnarray}
Using the metric Eq. \eqref{eq10} and connection Eq. \eqref{eq12}, we can get the equations of motion of $f(Q)$ theory for slowly rotating spherically symmetric metric for first order $\mathcal{O}(\epsilon)$,
\begin{align}\label{eq13}
    \kappa\mathcal{T}_{tt} =& \frac{e^{A-B}}{2r^{2}}\left\{r^{2}e^{B}f+2f_{Q}'r(e^{B}-1)\right. \nonumber\\
    &\left.+f_{Q}\left[(e^{B}-1)(2+rA')+(1+e^{B})rB'\right]\right\},\nonumber\\[5pt]
     \kappa\mathcal{T}_{rr} =& \frac{-1}{2r^{2}}\left\{r^{2}e^{B}f+2f_{Q}'r(e^{B}-1)\right.\nonumber\\
     &\left.+f_{Q}\left[(e^{B}-1)(2+rA'+rB')-2rA'\right]\right\},\nonumber\\[5pt]
     \kappa\mathcal{T}_{\theta\theta} =& -\frac{r}{4e^{B}}\left\{f_{Q}\left[-4A'-r(A')^{2}-2rA''+rA'B'\right.\right.\nonumber\\
     &\left.\left.+2e^{B}(A'+B')\right]+2e^{B}rf-2f_{Q}'rA'\right\},\nonumber\\[5pt]
     \kappa\mathcal{T}_{t\phi} =& \frac{1}{4}r\epsilon e^{-B}\sin^2(\theta)\Bigg\{2f_{Q}r\omega''-\left(f_{Q}\left(rA'+rB'-8\right)-2rf_{Q}'\right)\omega'\nonumber\\
     &+~\Big(2f_{Q}'\left(-rA'+e^{B}+1\right)+2\left(-f_{Q}rA''+f_{Q}e^{B}B'+fre^{B}\right)\nonumber\\
    &+f_{Q}A'\left(rB'+2e^{B}-4\right)-f_{Q}rA'^2\Big)~\omega\Bigg\}~.
\end{align}
where $Q = \frac{(e^{-B}-1)(A'+B')}{r}$ and $f_{Q}' = f_{QQ}\frac{dQ}{dr}$. As noticed from Eq. \eqref{eq13}, the field equations for the diagonal components remain unchanged. However, the inclusion of rotation introduces an additional field equation arising from the $t\phi$ component. In the slowly rotating case, the non-metricity tensor retains a similar form at $\mathcal{O}(\epsilon)$. Assuming the star is composed of a perfect fluid with a given EoS, the energy-momentum tensor is expressed as
\begin{eqnarray}\label{eq14}
    \mathcal{T}_{\mu\nu}= (\rho+p)u_{\mu}u_{\nu}+pg_{\mu\nu}
\end{eqnarray}
where the four-velocity components are
\begin{eqnarray*}
     u^r = u^{\theta} = 0, \quad u^{\phi} = \Omega u^t, \quad u^t = \Big[- (g^{tt} +
   2 \Omega g^{t \phi} + \Omega^2 g^{\phi \phi})\Big]^{1 / 2}
\end{eqnarray*}
The field equations Eq. \eqref{eq13} can then be written as a set of equations that include the TOV equations for $f(Q)$ gravity, describing the internal structure of neutron stars, as shown in \cite{Alwan_2024_2024_011} as
\begin{align}\label{toveq}
    A'' =& \frac{2 e^B \left( r (f(Q) + 2 p \kappa) + 
f_Q \left( A' + B' \right) \right)}{2f_Q r}\nonumber\\
&-\frac{A' \left( f_Q \left( 4 + r A' - r B' \right) + 2 f_{QQ} r Q' \right)}{2f_Q r}~,\nonumber\\
B' =& \frac{-\kappa e^B (p + \rho) r + f_Q A'}{f_Q}~, \nonumber\\
p' =& -\frac{(p+\rho)}{2}A'~.
\end{align}
An additional equation is obtained from the non-diagonal field equation that provides the angular velocity profile of the star as
\begin{eqnarray}\label{eq15}
    \bar{\omega}'' &=& f_2(A', B', Q', Q, r)\bar{\omega'}+f_1(A'', A', B', B, Q', Q, r, \rho)\bar{\omega}\nonumber\\
    &&+f_0(A'', A', B', B, Q', Q, r, \rho, p)~,\nonumber\\ \nonumber\\
    f_2 &=& \frac{f_{Q} r \left(A'+B'\right)-2rf_{Q}'-8 f_{Q}}{2 f_{Q} r}~, \nonumber\\[5pt]
    f_1 &=& \frac{2f_{Q}r A''-2 e^{B} \left(f_{Q} \left(A'+B'\right)+r (f(Q)- 16 \pi \rho )\right)}{2f_{Q} r} \nonumber\\
    &&+\frac{f_{Q}A' \left(r A'-r B'+4\right)-2f_{Q}' \left(-r A'+e^{B}+1\right)}{2f_{Q} r}~, \nonumber\\[5pt]
    f_0 &=& \frac{-2 f_Q r A''-f_Q \left(A'-B'\right) \left(r A'+2\right)}{2 f_Q r}\nonumber\\
    &&+\frac{2 e^B r (f(Q) -f_Q Q+16 \pi  p)+2 \left(e^B+1\right) f_{QQ} Q'}{2 f_Q r}~.
    \end{eqnarray}
Here, $\bar{\omega} := \Omega - \omega(r)$ represents the difference between the angular velocity $\Omega$ observed by a free-falling observer at infinity and the angular velocity of a fluid element measured by a local stationary observer within the fluid and $\Omega$ is spin frequency. In this context, we rescale $\bar{\omega}$ by defining the normalized metric function $\bar{\omega}/\Omega$ as a function of the radial coordinate, which makes the equations dimensionless.

From this, we can observe that the system of equations in Eq. \eqref{toveq} and Eq. \eqref{eq15} is separable, where the presence of the equation for $\bar{\omega}$ does not change Eq. \eqref{toveq}. This holds only at $\mathcal{O}(\epsilon)$. At $\mathcal{O}(\epsilon^2)$, this no longer applies, and the resulting equations become more complex. From Eq. \eqref{toveq} and Eq. \eqref{eq15}, we can obtain the solution of these TOV ODEs by integrating the equations from the center of the star outward. At the surface of the star, $\bar{\omega}$ has a boundary condition given by
\begin{equation}\label{eq16}
\bar{\omega} = 1 - \frac{2 J/\Omega}{r^3},
\end{equation}
where $J$ is the total angular momentum of the star, and the moment of inertia is given by $I = J/ \Omega$. In the exterior region of the star, Eq. \eqref{eq15} simplifies to
\begin{equation}\label{eq17}
\bar{\omega}'' - \frac{4\bar{\omega}'}{r} = 0, \end{equation}
which same with the form found in \cite{Hartle_1967_150_1005}. The solutions of these interior and exterior equations must match at the boundary condition Eq. \eqref{eq16}. Once this is done, we can extract the value of the moment of inertia, which is derived as
\begin{equation}\label{eq18}
I := \frac{J}{\Omega} = \frac{R^4}{6\Omega} \left(\frac{d\bar{\omega}}{d r}\right)_R,
\end{equation}
where $R$ is the radius of the star at the surface. Furthermore, at the surface of the star, the pressure approaches zero, allowing us to define the boundary value as  
\begin{equation}\label{pressure}
    p(R) \approx 0.
\end{equation}  
For the non-metricity, according to the formulation of covariant $f(Q)$, the condition $A' + B' = 0$ in vacuum leads to $Q(r) = 0$ outside the star. This can be readily obtained by evaluating Eq. \eqref{eq13}. Using the relations $A'(r) + B'(r) = 0$ and $e^{A(r)} = e^{-B(r)}$, which are easily derived from Eq. \eqref{eq13}, the exterior solution of the star can be written as:  
\begin{equation}\label{eqext}
e^{-B(r)} = e^{A(r)} = 1 + \frac{C}{r} + \frac{f(Q)|_{0}}{6f_Q|_0}r^2,
\end{equation}  
where $C$ is an integration constant. This solution resembles the Schwarzschild-de Sitter (SdS) solution with the cosmological constant $\Lambda = \frac{f(Q)|_{0}}{2f_Q|_0}$. The transition from the interior to the exterior of the star must satisfy the junction conditions discussed in our previous work (\cite{Alwan_2024_2024_011}). After determining the boundary values of the system, we require initial values for each variable to solve these ODEs. In this discussion, we use the following initial values:  
\begin{equation}\label{ivp}
B = 0, ~~~ A' = 0, ~~~ A = A_0, ~~~ Q = 0, ~~~ p = p_c, ~~~ \bar{\omega}' = 0, ~~~ \bar{\omega} = \bar{\omega}_c~,
\end{equation}  
where $\bar{\omega}_c$ and $A_0$ is a constant that can be chosen arbitrarily and will later be matched with the boundary condition at the star's surface using the shooting method. Meanwhile, $p_c$ is the pressure at the center of the star, which serves as a parameter when solving the ODEs using EoS.

To solve these ODEs, we also need to use dimensionless physical variables, defined as  
\begin{equation}\label{dimensionless}
\hat{r} = \frac{r}{r_g}, \quad \hat{p} = \frac{p}{p_g}, \quad \hat{\rho} = \frac{\rho}{\rho_g}, \quad \hat{Q} = Q r_g^2, \quad \hat{m} = \frac{m}{M_\odot},
\end{equation}  
where  
\begin{equation}\label{constant}
r_g = \frac{G M_\odot}{c^2}, \quad p_g = \frac{M_\odot c^2}{r_g^3}, \quad \rho_g = \frac{M_\odot}{r_g^3}.
\end{equation}  
In this context, $M_\odot$ represents the mass of the Sun, $c$ is the speed of light, and $G$ is the gravitational constant. These constants are expressed in the cgs unit system, where $M_\odot \approx 1.989 \times 10^{33} \, \mathrm{g}$, $c \approx 2.997 \times 10^{10} \, \mathrm{cm/s}$, and $G \approx 6.674 \times 10^{-8} \, \mathrm{dyne \, cm^2/g^2}$. Note that $\bar{\omega}$ is not rescaled, as it is already normalized by $\Omega$. With this setup, we can compute the profiles of the neutron star, particularly $\bar{\omega}$ and the moment of inertia as properties of the NS, which will be discussed further in this paper. Using some realistic EoS, we apply the Runge-Kutta method available in the \texttt{scipy.integrate.solve\_ivp} package (\cite{2020SciPy-NMeth}) in Python to numerically integrate the ODEs and obtain the solutions. This package provides a flexible and computationally efficient approach for solving initial value problems in ODEs.

\section{Impact of $f(Q)$ models on Slowly Rotating case}\label{Sec3}
Using the TOV equations and the numerical setup discussed in the previous section, we calculate the profile of $\bar{\omega}$ and the properties of the neutron star that can be obtained in the slowly rotating case, specifically the moment of inertia. From the results, we can compute the universal $\bar{I}-C$ relation and demonstrate the deviations of various $f(Q)$ models from this relation, independent of the EoS. In this section, we analyze the rotation effects in the extended $f(Q)$ gravity using several models, specifically the linear, quadratic, exponential, and logarithmic models, given by,
\begin{eqnarray}
    f(Q) &=& \alpha_1 Q + \beta_1 ~, \quad\quad f(Q) = Q + \alpha_2 Q^2~,\nonumber\\
    f(Q) &=& Q + \alpha_3 e^{\beta_3 Q}~, \quad f(Q) = Q - \alpha_4 \ln(1-\beta_4 Q)~,
\end{eqnarray}
where $\alpha$ and $\beta$ are model parameters. The non-linear $f(Q)$ models have been extensively explored in our previous work (\cite{Alwan_2024_2024_011}). Additionally, this paper examines the linear $f(Q)$ model, which has been widely studied in the context of compact star scenarios (\cite{Lohakare_2023_526_3796, Bhar_2024_43_101391, Chaudharya_2024_09_049, Gul_2024_84_8, Errehymy:2024qpu}). However, the linear model in this case is slightly different because we consider an isotropic fluid, not anisotropic, as the matter inside the neutron star. By using the assumption of the contracted energy-momentum conservation $\nabla^{\mu}\mathcal{T}_{\mu\nu} = 0$, we obtain the condition where $f_{QQ} = 0$. In this model, $\alpha_1$ is a constant of integration from the $f_{QQ} = 0$ condition, and $\beta_1$ is a constant of integration from $f_Q$, resulting in the linear form $f(Q) = \alpha_{1} Q + \beta_{1}$.

\subsection{Moment of Inertia}
\begin{figure*}
    \centering
    \begin{subfigure}[b]{0.50\textwidth}
        \centering
        \includegraphics[width=80mm]{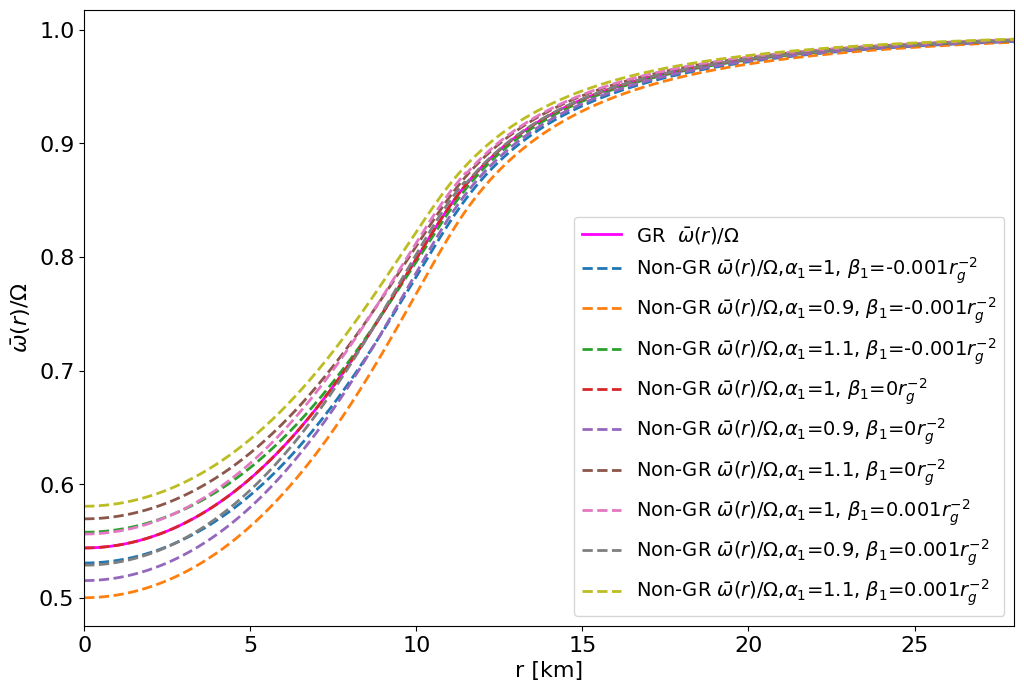}
        \caption{$\bar{\omega}/\Omega$ profile of $f(Q)=\alpha_{1} Q+\beta_{1}$}
        \label{fig:wlin}
    \end{subfigure}%
    \hfill
    \begin{subfigure}[b]{0.50\textwidth}
        \centering
        \includegraphics[width=80mm]{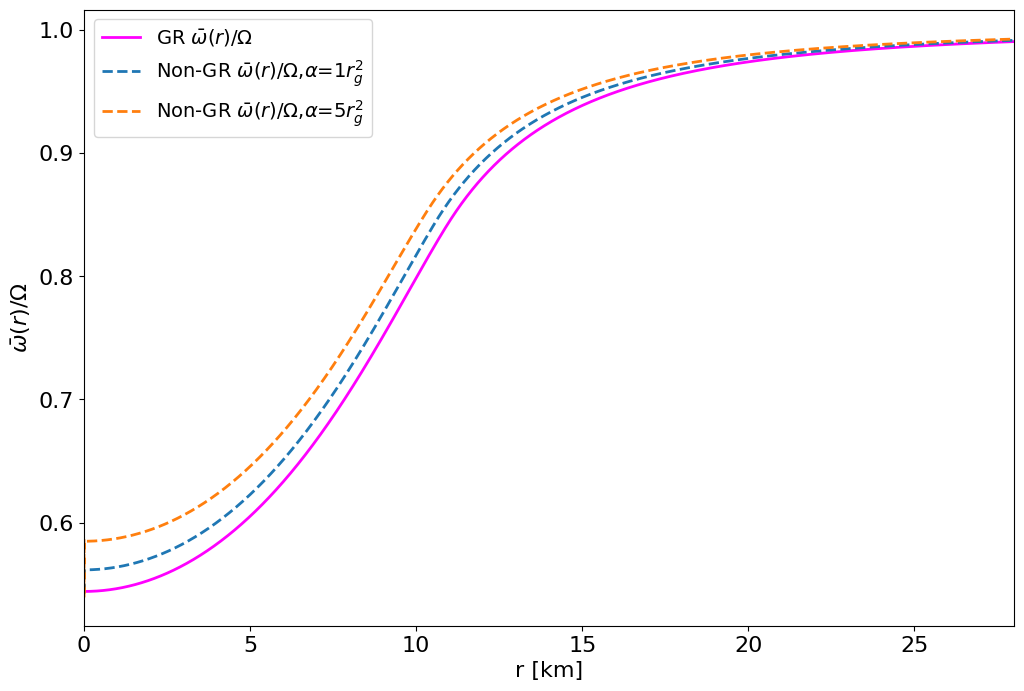}
        \caption{$\bar{\omega}/\Omega$ profile of $f(Q)=Q+\alpha_{2} Q^2$}
        \label{fig:wquad}
    \end{subfigure}%
    \vskip\baselineskip
    \begin{subfigure}[b]{0.50\textwidth}
        \centering
        \includegraphics[width=80mm]{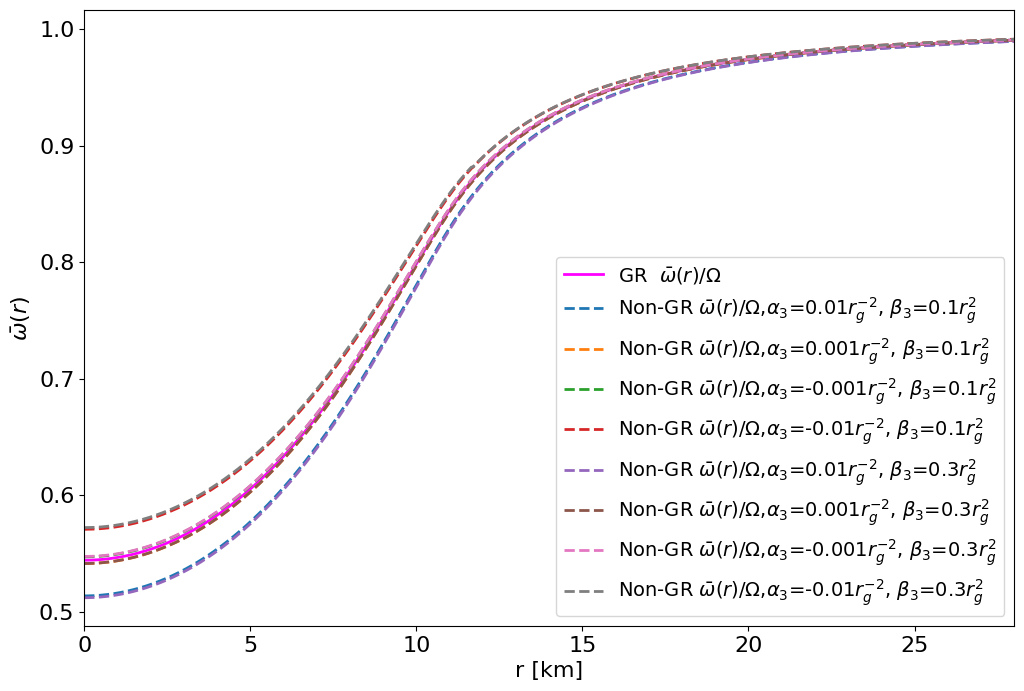}
        \caption{$\bar{\omega}/\Omega$ profile of $f(Q) = Q + \alpha_{3} e^{\beta_{3} Q}$}
        \label{fig:wexp}
    \end{subfigure}%
    \hfill
    \begin{subfigure}[b]{0.50\textwidth}
        \centering
        \includegraphics[width=80mm]{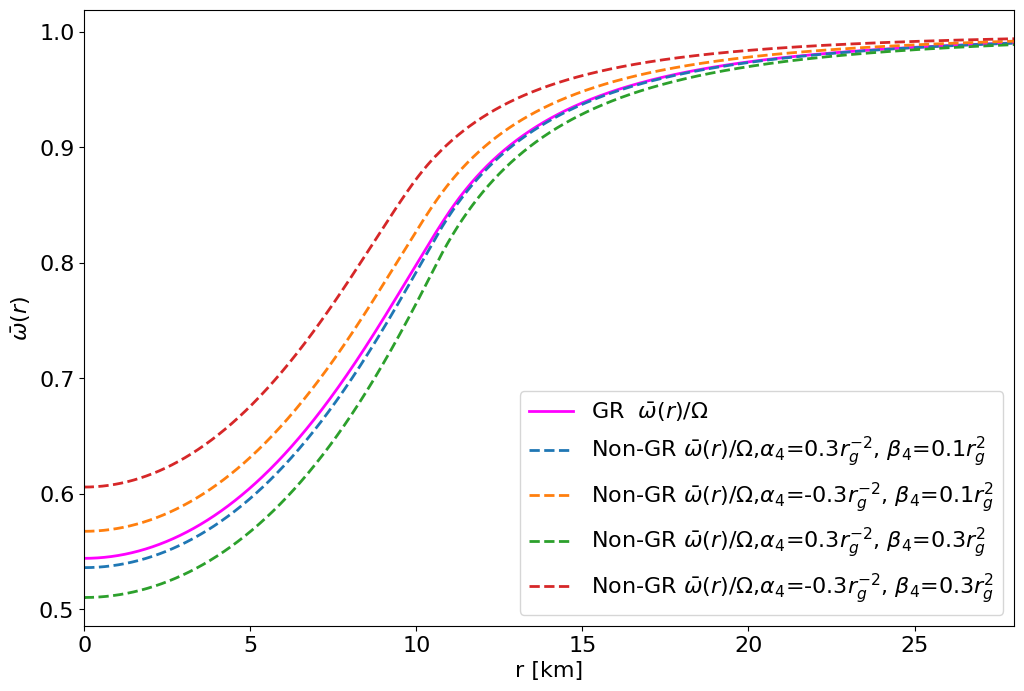}
        \caption{$\bar{\omega}/\Omega$ profile of $f(Q) = Q - \alpha_{4} \ln(1-\beta_{4} Q)$}
        \label{fig:wlog}
    \end{subfigure}
    \caption{The normalized $\bar{\omega}/\Omega$ profile as a function of radius, computed using the SLy4 EoS at a density of $\rho = 1 \times 10^{15}$ g/cm$^3$. For each model, various values of $\alpha$ and $\beta$ are considered. By applying the matching condition between the interior and exterior solutions, different initial values are determined for each parameter, highlighting the deviations of each $f(Q)$ model from GR. In this graph, the star surface radius in GR is approximately $11.7$ km.}
    \label{fig:omega}
\end{figure*}

Utilizing the numerical framework outlined in Sec. \ref{Sec2B}, we calculate the $\bar{\omega}$ profiles for each model, as illustrated in Fig. \ref{fig:omega}. This figure presents $\bar{\omega}$ normalized by $\Omega$ for different valufor each model, employing the SLy EoS (\cite{Douchin_2001_380_151}). To determine the surface of the star, we set the condition $p(R) \leq 10^{-8} p_c$. At $\rho = 1 \times 10^{15}$, the models exhibit noticeable deviations from GR. Each model utilizes distinct initial values of $\bar{\omega}$, which are calibrated through the shooting method to satisfy the exterior boundary conditions specified at the surface of the star, as given in Eq. \eqref{eq17}. This indicates that the deviations become stronger toward the center of the star, potentially due to the influence of non-metricity. As one moves outward from the origin, the deviations from the central conditions gradually decrease, approaching the asymptotic condition at infinity, where the ratio $\bar{\omega}/\Omega$ converges to $1$. As anticipated from our prior analysis, smaller values of $\rho_c$ lead to smaller deviations from GR, which become almost negligible. Conversely, larger values of $\rho_c$ cause stronger deviations from GR, aligning with the observed characteristics of neutron star configurations within the framework of covariant $f(Q)$ gravity (\cite{Alwan_2024_2024_011}). Changing the parameters $\alpha$ and $\beta$ can effectively reduce the deviations in the central values of $\bar{\omega}$. By decreasing these parameter values, the results can approach or even coincide with the GR. This is evident in Fig. \ref{fig:wlin}, where for $\alpha_{1} = 1$ and $\beta_{1} = 0$, the graph aligns perfectly with GR, indicating that the TOV equations revert to the GR case.

One of the concerns with the $\bar{\omega}$ profiles is the behavior of the exponential model. In this model, we only use values of $\alpha_3 = \pm 0.001$ and $\pm 0.01$, since larger deviations in $\bar{\omega}$ are observed when $\alpha_3 = 0.1$. For example, when $\alpha_3 = \pm 0.01$, the difference in $\bar{\omega}_c$ compared to GR ranges from 5.2\% to 5.9\%. However, at $\alpha_3 = \pm 0.1$, the deviations become much more pronounced, reaching 32\% for $\alpha_3 = -0.1$ and 210\% for $\alpha_3 = 0.1$. These results suggest that the rotational effects in the exponential model can lead to substantial deviations, potentially making neutron stars less stable. This is further supported by the $\bar{I}-C$ relation, which is discussed later. This result differs from our previous work, where values of $\alpha_3$ ranging from $0.1$ to $1$ produced noticeable yet stable deviations in the mass-radius relation, indicating a potential limitation of the exponential model for slowly rotating stars. Additionally, the parameter $\beta_3$ in this model seems to have minimal effects on the $\bar{\omega}$ profile. In earlier studies, values of $\alpha_3 = 0.1$ with $\beta_3 = 0.1$ and $\beta_3 = 0.5$ led to a maximum mass difference of up to 6.8\%. This discrepancy might arise because the value of $\alpha$ in the current analysis is too small, making the fine-tuning effect of $\beta$ negligible. For other models, such as linear, quadratic, and logarithmic, the differences in $\bar{\omega}_c$ range from 2.7\% to 12.3\%. Although the logarithmic model, for example, shows differences as large as 12.3\%, the resulting neutron stars remain stable and exhibit similar behavior when the parameters $\alpha_4$ and $\beta_4$ are varied, consistent with our previous work. In the linear model of $f(Q)$, where $\beta_1$ serves as the integration constant, $\beta_1$ proves to be highly sensitive, as it affects the magnitude of $f(Q)$, which in turn impacts the stability of the star. Thus, in our numerical calculations, we used very small values for $\beta_1$. On the other hand, $\alpha_1$ is more flexible as it scales $Q$.

\begin{figure*}
    \centering
    \begin{subfigure}[b]{0.50\textwidth}
        \centering
        \includegraphics[width=80mm]{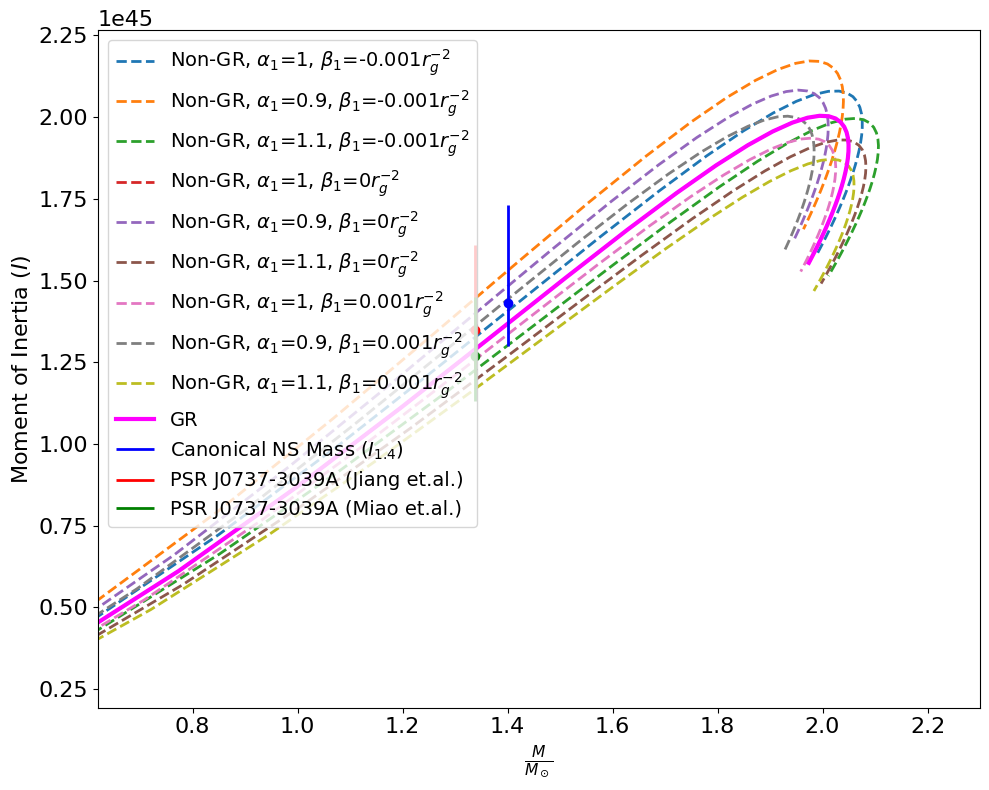}
        \caption{Moment of inertia of $f(Q)=\alpha_{1} Q+\beta_{1}$}
        \label{fig:moilin}
    \end{subfigure}%
    \hfill
    \begin{subfigure}[b]{0.50\textwidth}
        \centering
        \includegraphics[width=80mm]{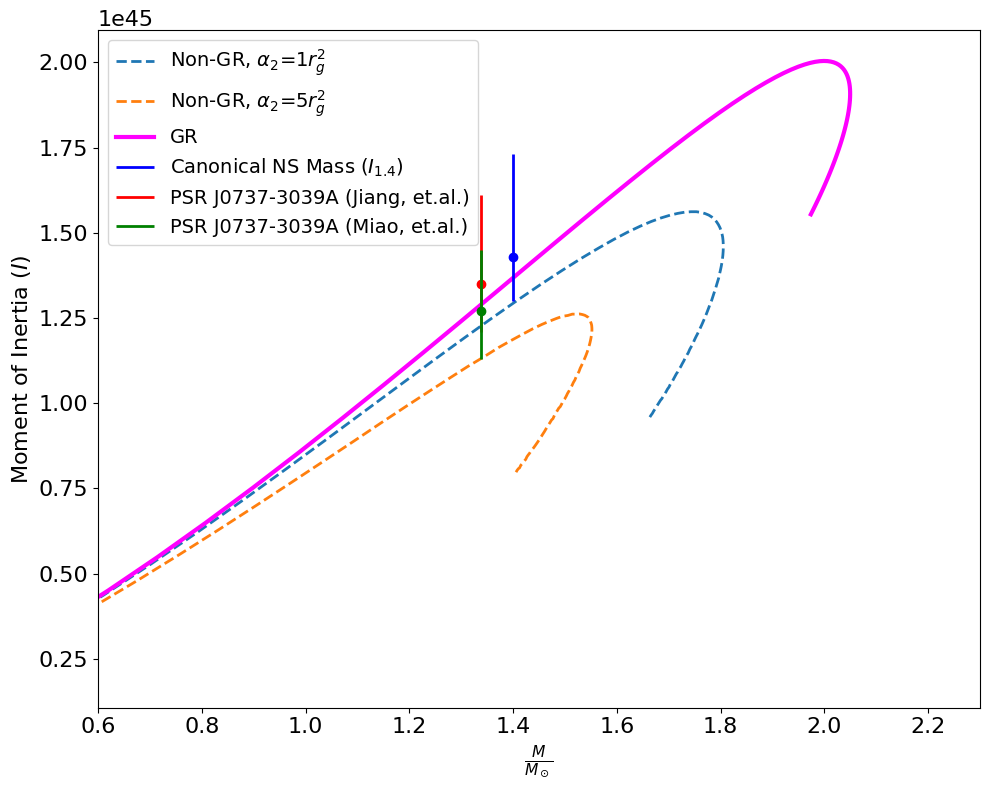}
        \caption{Moment of inertia of $f(Q)=Q+\alpha_{2} Q^2$}
        \label{fig:moiquad}
    \end{subfigure}%
    \vskip\baselineskip
    \begin{subfigure}[b]{0.50\textwidth}
        \centering
        \includegraphics[width=80mm]{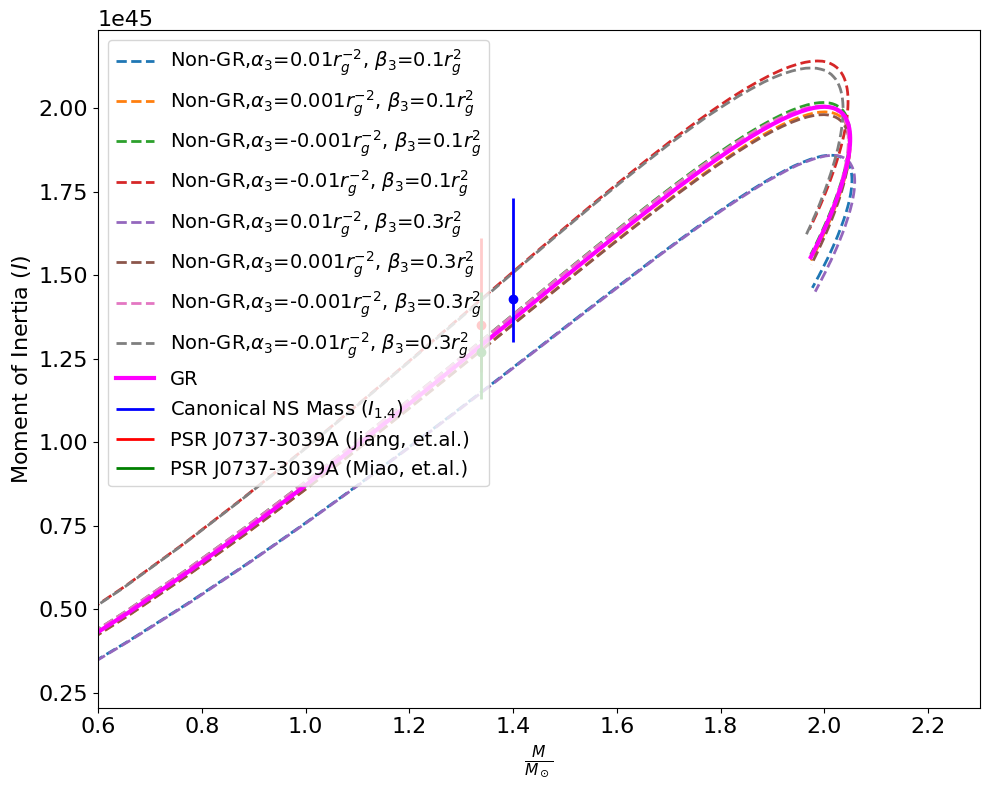}
        \caption{Moment of inertia of $f(Q) = Q + \alpha_{3} e^{\beta_{3} Q}$}
        \label{fig:moiexp}
    \end{subfigure}%
    \hfill
    \begin{subfigure}[b]{0.50\textwidth}
        \centering
        \includegraphics[width=80mm]{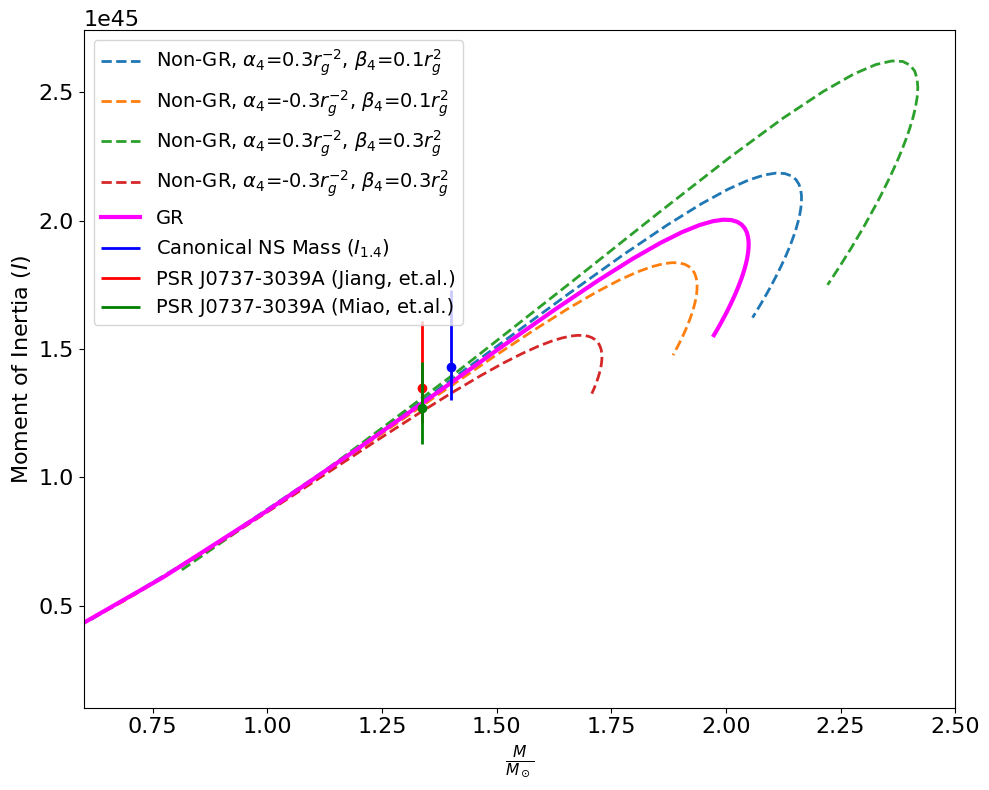}
        \caption{Moment of inertia of $f(Q) = Q - \alpha_{4} \ln(1-\beta_{4} Q)$}
        \label{fig:moilog}
    \end{subfigure}
    \caption{Moment of inertia as a function of mass for SLy EoS using different values of the parameter $\alpha$ and $\beta$ of each model, in units of $\text{g cm}^2$. The red and green vertical lines represent the moment of inertia constraints for PSR J0737-3039A ($M = 1.338~ M_\odot$) from \citet{Jiang:2020uvb} and \citet{Miao_2021_515_5071}, respectively, derived using different methods at $90\%$ Credible Interval. The blue vertical line shows the constraint for a canonical neutron star mass at $M = 1.4~ M_\odot$ from \citet{Jiang:2020uvb} calculation. }
    \label{fig:moiall}
\end{figure*}

After calculating the $\bar{\omega}$ profiles for each model, we proceed to compute the moment of inertia using the values of $\bar{\omega}$ at the surface, as given by Eq. \eqref{eq18}. In Fig. \ref{fig:moiall}, we plot the moment of inertia of the neutron star as a function of its mass for each model, with different parameter values. The results are consistent with our previous calculations of $\bar{\omega}$, where, as the parameters approach zero, the moment of inertia converges to the value predicted by GR. If we observe the moment of inertia of all models at low masses, the moment of inertia of the neutron star closely matches the GR result, except for the linear and the exponential model. In the models, the largest deviation reaches $\sim 20\%$ at low mass. This reinforces the presumption that the rotational effects significantly affect the stability of neutron stars obtained in the exponential model. Although other references, such as \cite{Staykov_2014_2014_006}, show that the moment of inertia exhibits larger deviations than the maximum mass deviation, in this model, the deviations are so large that we can only calculate for small values of $\alpha$ up to the order of $\mathcal{O}(10^{-2})$ for exponential model and $\mathcal{O}(10^{-3})$ for linear model. Another possible explanation is that if we expand the exponential term using a Taylor series, we can obtain $f(Q) = Q + \alpha_3(1 + \beta_3 Q + \frac{\beta_3^2 Q^2}{2} + \mathcal{O}(Q^3))$, where $\alpha_3$ and $\beta_3$ represent terms that are part of the exponential expansion. There is a similarity with the linear model $f(Q)$, where the term $\alpha_3$ in this case plays the role of $\beta_1$ in linear $f(Q)$, and the term $\alpha_3 \beta_3 Q$ will become an additional linear $Q$-term, similar to $\alpha_1 Q$ in the linear $f(Q)$. The quadratic term becomes so small that it can be neglected. If we compare $\alpha_3$ and $\beta_1$, we can observe that both models only achieve stable moments of inertia at small maximum mass deviations, particularly at orders around $10^{-3} r_g^{-2}$. This suggests that for the slowly rotating case, both models can only stabilize neutron stars when the maximum mass deviation is small, as can be seen in Fig. \ref{fig:moilin} and \ref{fig:moiexp}. For the quadratic model, the maximum deviation occurs at $\alpha_2 = 5r_g^2$, with a deviation of approximately $\sim 49\%$. We tested various values of $\alpha_2 = \{10, 50, 10^2, 10^3, 10^4\}r_g^2$ and observed that the maximum deviation in the moment of inertia decreases with $\alpha_2>5r_g^2$. We also found that the moment of inertia tends to stagnate at $\alpha_2 = 10^2r_g^2$, which is lower than the corresponding parameter value in quadratic $f(R)=R+\gamma R^2$ gravity, where stagnation occurs at $\gamma=10^4r_g^2$ (\cite{Staykov_2014_2014_006}). Unfortunately, there are currently no observational constraints on the quadratic $f(Q)$ parameter. In contrast, $f(R)$ gravity has a constraint of $\gamma \gtrsim 2.3 \times 10^5$ (\cite{Joachim_2010_81_104003}), derived from the Gravity Probe B experiment. The logarithmic models exhibit similar behavior, where the deviations exceed the maximum mass deviation. However, they demonstrate greater stability compared to the exponential model, particularly for parameters that achieve significantly higher maximum masses.

Constraining gravity theories using the moment of inertia is not a straightforward task. Unlike pulsar timing measurements that can tightly constrain the mass, or X-ray observations that provide information about the radius through pulse-profile modeling, the moment of inertia requires high-quality data over long observational timescales and cannot yet be determined with the same accuracy. One of the pulsars for which the moment of inertia has been estimated is PSR J0737-3039A, with a precisely measured mass of $M = 1.338~M_\odot$. For example, \cite{Landry:2018jyg} estimated its moment of inertia as $I_A \sim 1.15^{+0.38}_{-0.24} \times 10^{45}~\text{g}~\text{cm}^2$ for $M=1.338~M\odot$, using tidal measurements from GW170817. Later, \cite{Silva:2020acr} predicted the moment of inertia as $I_A \sim 1.68^{+0.53}_{-0.48} \times 10^{45}~\text{g}~\text{cm}^2$ at the same mass, based on NICER mass–radius measurements of PSR J0030+0451. In this paper, we also compare our results with the constraints obtained by \cite{Jiang:2020uvb} and \cite{Miao_2021_515_5071}. In Jiang’s analysis, they first estimated $I$ for a canonical neutron star of $1.4~M\odot$ and then extrapolated it to the mass of PSR J0737-3039A, obtaining $I_A \sim 1.35^{+0.26}_{-0.14} \times 10^{45}~\text{g}~\text{cm}^2$. In contrast, \cite{Miao_2021_515_5071} directly calculated the moment of inertia at $M = 1.338~M\odot$ using an explicit hadron–quark phase transition model with the constant speed of sound (CSS) parameterization, combining GW170817, GW190425, and NICER data (PSR J0030+0451 and PSR J0740+6620).
In all four models $f(Q)$ that we have studied, the predicted moment of inertia values are consistent with the ranges reported in these works. However, certain parameter values that result in a lower moment of inertia do not satisfy the prediction given by~\cite{Jiang:2020uvb} for canonical NS mass. Unfortunately, the moment of inertia-mass relation still strongly depends on the EoS, making it insufficient for constraining gravity models. Therefore, a universal relation is required to establish an EoS-independent framework, which will be further discussed in the next section.

\subsection{Universal $\bar{I}-C$ Relation}
After obtaining the characteristics of the neutron star in the slowly rotating case, including the moment of inertia, we can calculate one of the so-called universal relations, which provides a connection between stellar properties that is nearly independent of the underlying EoS. Among the first such relations to be discovered was the link between the normalized moment of inertia and compactness, $I/MR^2$–$C=M/R$, proposed by \citet{Ravenhall_1994_424_846} and later refined by \citet{Lattimer:2000nx}. This relation was further extended by \citet{Breu_2016_459_646}, who demonstrated its quasi-universal nature across a wide range of rotation rates and EoS in GR. The universal $\bar{I}$–$C$ relation can be expressed as (\citet{Breu_2016_459_646})
\begin{equation}\label{i-cb}
\bar{I} = a_1 C^{-1} + a_2 C^{-2} + a_3 C^{-3} + a_4 C^{-4},
\end{equation}
where the moment of inertia is normalized as $\bar{I} = I c^2/(GM^3)$ and compactness as $C = GM/R$, with $a_{1,\dots,4}$ denoting the fitting coefficients. This relation has been revisited and refined in several subsequent studies. Most recently, \cite{Suleiman:2024ztn} conducted an exhaustive analysis of the I–C relation using large ensembles of EoS models, including both nuclear-physics-based and agnostic EoS. Their use of agnostic ensembles allowed them to capture the residual EoS variability of the quasi-universal relation.

Beyond GR, the $\bar{I}$–$C$ relation has also been explored as a probe of alternative theories of gravity. For example, it has been applied in $f(R)$ gravity and scalar–tensor~(\cite{Staykov2016}), in scalar–torsion theories~(\cite{Boumaza:2021fns,Kehal:2023rhc}), and in the hordenski gravity theories~(\cite{Sakstein_2017_95_064013}). These studies demonstrate that the quasi-universality of the $\bar{I}$–$C$ relation extends beyond GR, although with potential deviations arising from the underlying geometric modifications. In this paper, we extend such analyses to $f(Q)$ gravity, examining how nonmetricity-induced corrections influence the behavior of this relation.

In our previous analysis, we employed four representative EoS—SLy, FPS, AP4, and MS1b—to compute the interior profiles and mass–radius relations of neutron stars. In the present study, we extend this framework by considering a substantially broader and more diverse set of EoS that encompass both relativistic mean-field (RMF)~(\cite{PhysRevC.95.065803_Hyperonic1}) and Skyrme–Hartree–Fock (SHF)~\cite{PhysRevC.85.035201_SMF}) formalisms. All selected models are constrained by empirical nuclear data, including finite-nuclei properties and the saturation characteristics of bulk nuclear matter, ensuring their reliability from microscopic to astrophysical scales~(\cite{Fortin2016, Kumar:2017xdy, Kumar2018_IOPB}). Here is the EoS list that we use in this paper:
\begin{itemize}
\item Nucleonic RMF:
        \begin{itemize}
            \item BKA20~(\cite{Agrawal2010_BKA20}), BSP~(\cite{Agrawal:2012rx}), IOPB-1~(\cite{Kumar2018_IOPB}), Model1~(\cite{Mondal:2015tfa_Model1}),
            \item MPA1~(\cite{Muther_1987_199_469}), MS1b~(\cite{Mueller_1996_606_508}), SINPA~(\cite{Mondal:2016roo_SINPA}),
            \item GM1~(\cite{PhysRevLett.67.2414_GM1}), G3~(\cite{PhysRevC.95.015801_G3}), 
            \item NL3~(\cite{PhysRevC.55.540_NL3}), and NL3$\omega$~(\cite{PhysRevLett.86.5647_NL3w}).
            \item Density-dependent (DD) variants:  
        DD2~(\cite{PhysRevC.81.015803_DD2}) and DDME2~(\cite{Gaitanos:2003zg_DDME2}).  
        In these models, the couplings $g_i(\rho_B)$ depend explicitly on the baryon density, improving the description of the symmetry energy at supranuclear densities.  
        \end{itemize}
    \item Hyperonic RMF extensions~(\cite{PhysRevD.101.034017_Hyperonic2, PhysRevC.95.065803_Hyperonic1}):  
        \begin{itemize}
            \item BSR2Y,BSR6Y
            \item DD2Y, and DDME2Y
            \item GM1Y
            \item NL3Y, NL3$\omega\rho$Y, NL3$\omega\rho$Yss, NL3Yss
        \end{itemize}
    
\item Skyrme–Hartree–Fock (SHF) models:  
    \begin{itemize}
        \item BSk20–BSk26~(\cite{Goriely:2010bm_BSk2, Potekhin:2013qqa_BSK23, Pearson:2018tkr_BSK25})  
        \item KDE0v1~(\cite{Agrawal:2005ix_KDE})  
        \item Rs~(\cite{Friedrich:1986zza_RS})  
        \item SK255, SK272~(\cite{Agrawal:2003xb_Sk255272})  
        \item SKa and SKb~(\cite{Kohler:1976fgx_SKab})  
        \item SkI2–SkI6~(\cite{Reinhard:1995zz_SkI25, PhysRevC.53.740_SkI6})  
        \item SkMP~(\cite{1989PhRvC..40.2834B_Skmp})  
        \item SKOp~(\cite{Reinhard:1999ut_SKop})  
        \item SLy230a~(\cite{Chabanat:1997qh_SLy230a}) and SLy2, SLy4, SLy9~(\cite{Douchin_2001_380_151})
    \end{itemize}
\end{itemize}
We also consider the microscopic BCPM model~(\cite{Sharma:2015bna_BCPM}), which is derived from the Argonne $v_{18}$ potential complemented by Urbana three-body forces. The complete list of the EoS used in this work is summarized in Table~\ref{tab:eos_details}.

Many of these EoS are unified, providing a consistent pressure–density relation from the crust to the core, while others represent core-only models supplemented by a SLY4 crust at low densities. All the selected EoS are thermodynamically stable. In terms of causality, the RMF models are inherently causal, while the SHF models are causal up to the central densities of the maximum mass configurations, except for the BSk20 and BSk26 models, which are marginally acausal at these densities~(\cite{Malik_2018}). Among the 53 EoS considered, 45 support a maximum mass exceeding the observational lower limit of $2.08 \pm 0.07 \, M_\odot$~(\cite{Fonseca:2021wxt}), while the remaining eight yield maximum masses in the range $1.93$–$1.99 \, M_\odot$ within GR~(\cite{Demorest:2010bx, Antoniadis:2013pzd}). These softer EoS are retained to examine whether the modified gravity framework can effectively stiffen the stellar configuration at high densities. For instance, the softest EoS in our sample, BKA20, which predicts a maximum mass of $1.952\,M_\odot$ in GR, reaches $2.094\,M_\odot$ in the logarithmic $f(Q)$ model with $\alpha = 0.3\,r_g^{-2}$ and $\beta = 0.1\,r_g^{2}$. This behavior highlights how geometric modifications can extend the viable EoS parameter space while maintaining consistency with current astrophysical constraints. The diversity in microphysical assumptions, particle content, and stiffness—from soft models such as BCPM and BSP to stiff models like MS1b and NL3—enables a systematic exploration of the universality of the $\bar{I}$–$C$ relation across a broad range of nuclear models, stiffness levels, and stellar compositions, including hyperonic configurations. For the computation, we use a piecewise polytropic representation of the EoS, a phenomenological approach that offers a practical description of realistic nuclear models across different density regions.This approach, originally proposed by \cite{Read:2008iy}, further developed by \cite{Kumar:2019xgp}, and later refined using unified EoS by \cite{Suleiman:2022egw}, allows us to reproduce the macroscopic behavior of neutron-star matter across different density ranges without relying on the full tabulated data. The fitted parameters from these representations are then used consistently in our numerical calculations of stellar structure.

\begin{table*}
\caption{\label{tab:eos_details}Classification of the EoS used in this work and their representative colors in the plots.}
\renewcommand{\arraystretch}{1.3}
\setlength{\tabcolsep}{8pt}
\centering
\begin{tabular}{l c}
\hline\hline
\textbf{EoS Model} & \textbf{Color} \\ 
\hline
\multicolumn{2}{l}{\textbf{RMF (Nucleonic) Models}} \\
\hline
BKA20 & \cellcolor{magenta!80} \\
BSP & \cellcolor{sienna!80} \\
G3,GM1 & \cellcolor{cyan!80} \\
MPA1 & \cellcolor{dodgerblue!80} \\
MS1b & \cellcolor{darkviolet!80} \\
IOPB & \cellcolor{gold!80} \\
NL3, NL3$\omega\rho$ & \cellcolor{peru!80} \\
FSU, FSUGarnet & \cellcolor{teal!80} \\
BSR2, BSR6 & \cellcolor{blueviolet!80} \\
DD2, DDME2, DDHd & \cellcolor{limegreen!80} \\
TM1 & \cellcolor{crimson!80} \\
\hline
\multicolumn{2}{l}{\textbf{RMF (Hyperonic) Models}} \\
\hline
BSR2Y, BSR6Y & \cellcolor{blueviolet!80} \\
DD2Y & \cellcolor{limegreen!80} \\
DDME2Y & \cellcolor{springgreen!80} \\
GM1Y & \cellcolor{red!80} \\
NL3$\omega\rho$Y, NL3$\omega\rho$Yss, NL3Y, NL3Yss & \cellcolor{peru!80} \\
\hline
\multicolumn{2}{l}{\textbf{SHF (Nucleonic) Models}} \\
\hline
BSk20--BSk26 & \cellcolor{deeppink!80} \\
KDE0v1 & \cellcolor{slategray!80} \\
SK255, SK272, SKa, SKb, SKI2--SKI6, SkMP, SKOp & \cellcolor{deepskyblue!80} \\
SLY230a, SLY2, SLY4, SLY9 & \cellcolor{royalblue!80} \\
Rs & \cellcolor{yellowgreen!80} \\
SINPA & \cellcolor{olive!80} \\
Model1 & \cellcolor{hotpink!80} \\
\hline
\multicolumn{2}{l}{\textbf{(Nucleonic) \textit{a-b Initio} Model}} \\
\hline
BCPM & \cellcolor{orangered!80} \\
\hline\hline
\end{tabular}
\end{table*}

\begin{figure*}
    \centering
    \begin{subfigure}[b]{0.50\textwidth}
        \centering
        \includegraphics[width=80mm]{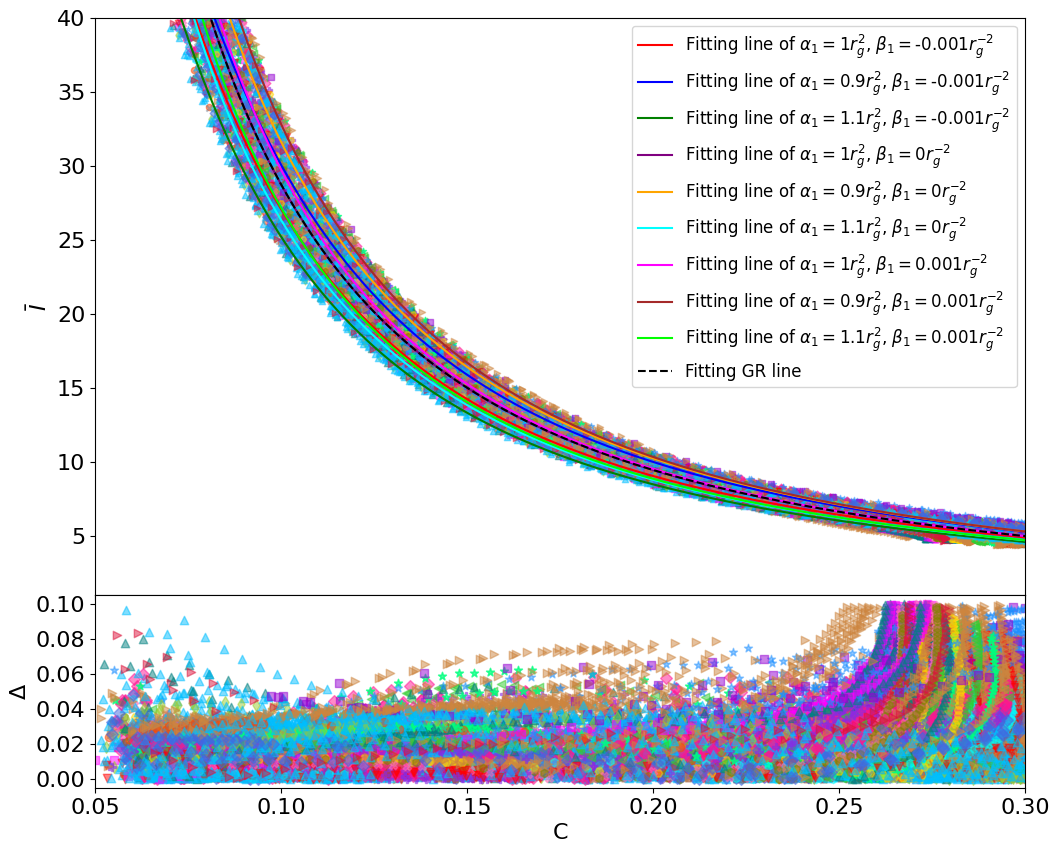}
        \caption{$\bar{I}-C$ relation of $f(Q)=\alpha_{1} Q+\beta_{1}$}
        \label{fig:fitlin}
    \end{subfigure}%
    \hfill
    \begin{subfigure}[b]{0.50\textwidth}
        \centering
        \includegraphics[width=80mm]{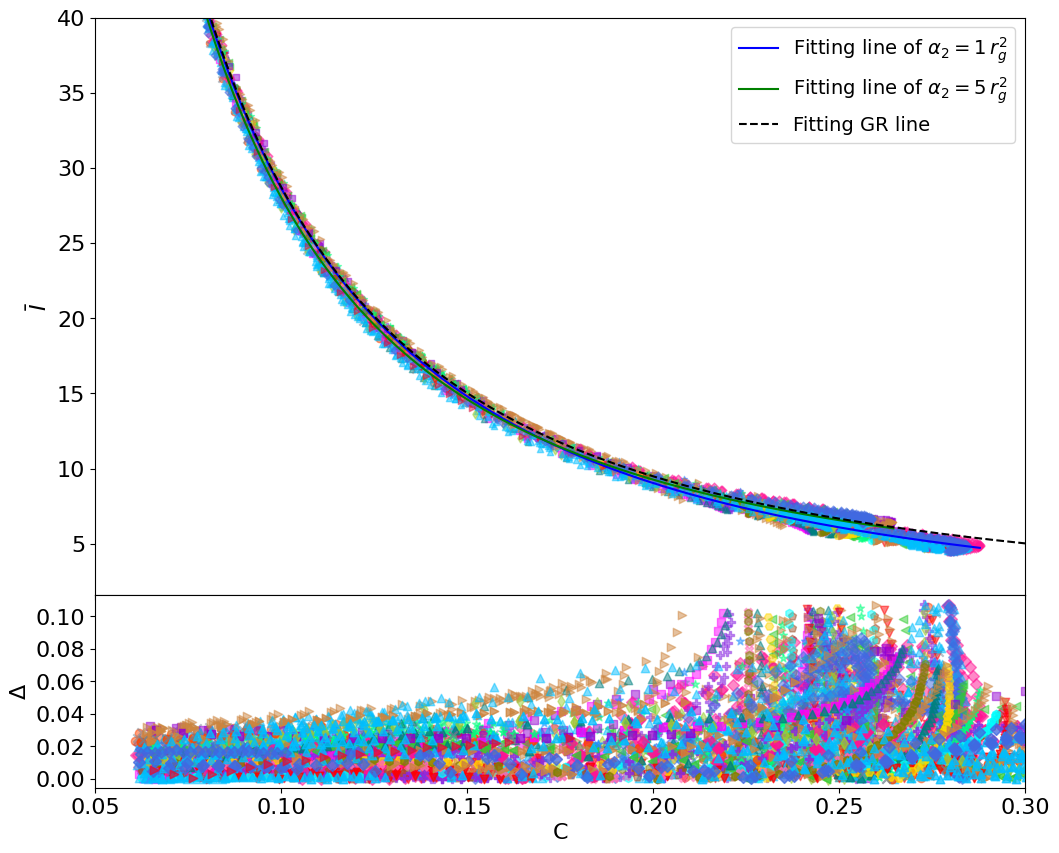}
        \caption{$\bar{I}-C$ relation of $f(Q)=Q+\alpha_{2} Q^2$}
        \label{fig:fitquad}
    \end{subfigure}%
    \vskip\baselineskip
    \begin{subfigure}[b]{0.50\textwidth}
        \centering
        \includegraphics[width=80mm]{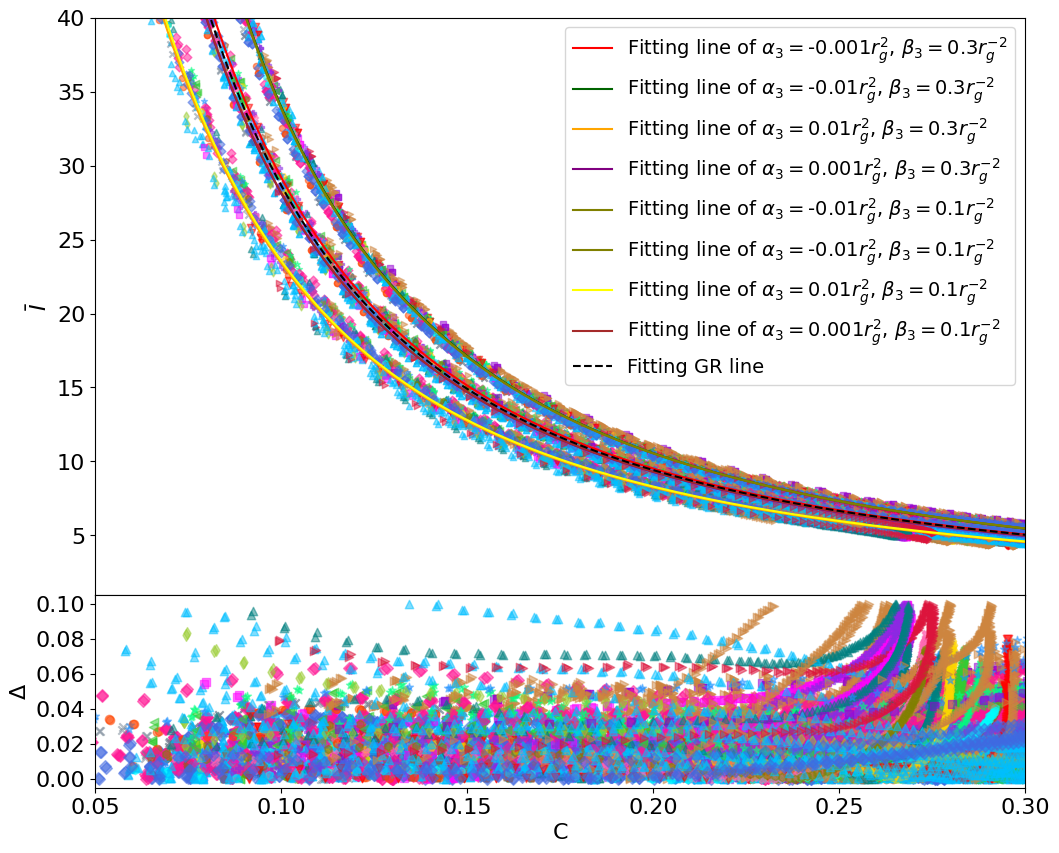}
        \caption{$\bar{I}-C$ relation of $f(Q) = Q + \alpha_{3} e^{\beta_{3} Q}$}
        \label{fig:fitexp}
    \end{subfigure}%
    \hfill
    \begin{subfigure}[b]{0.50\textwidth}
        \centering
        \includegraphics[width=80mm]{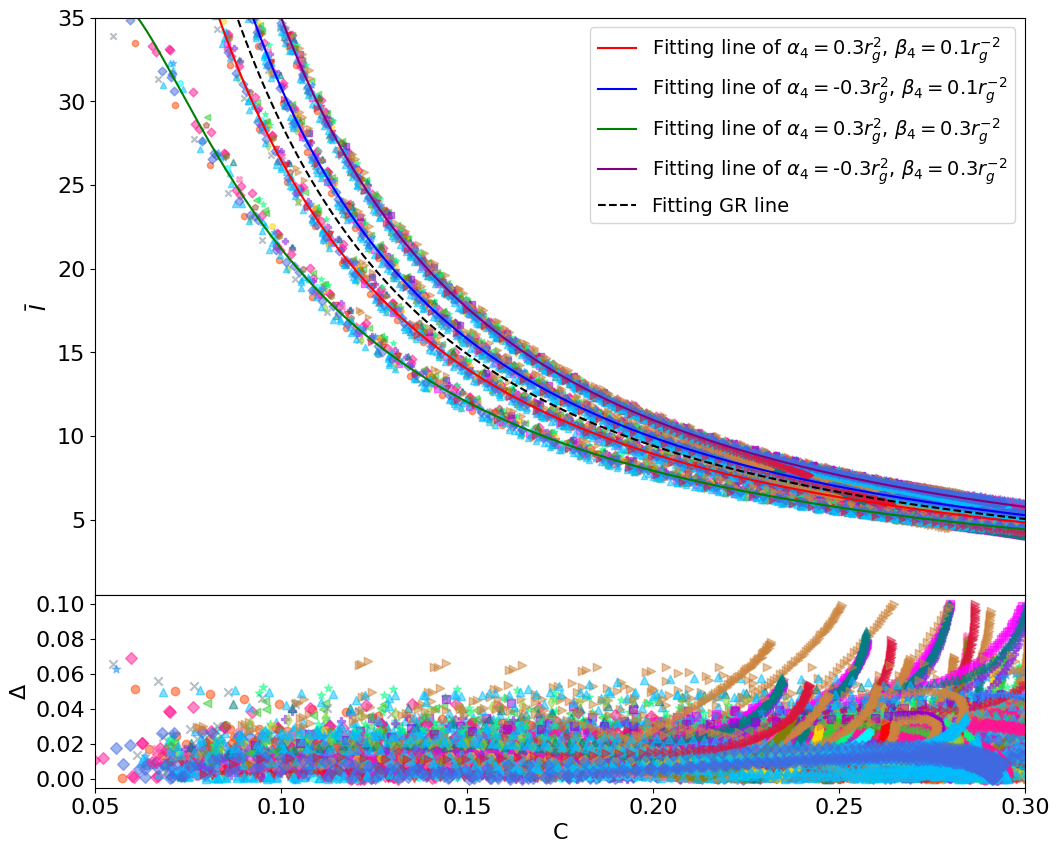}
        \caption{$\bar{I}-C$ relation of $f(Q) = Q - \alpha_{4} \ln(1-\beta_{4} Q)$}
        \label{fig:fitlog}
    \end{subfigure}
    \caption{Universal $\bar{I}-C$ relation of various $f(Q)$ model.}
    \label{fig:fitline}
\end{figure*}

\begin{figure*}
    \centering
    \begin{subfigure}[b]{0.50\textwidth}
        \centering
        \includegraphics[width=80mm]{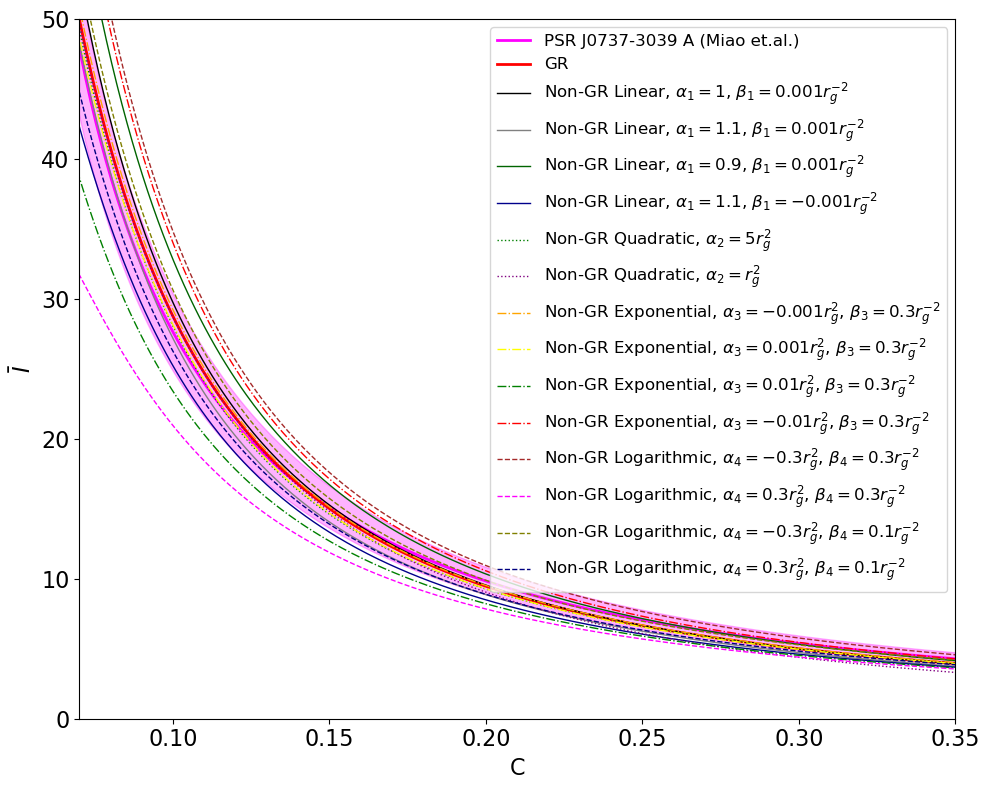}
        \caption{The best-fitting $\bar{I}-C$ relation from Table \ref{tab:fitting_coefficients}.}
        \label{fig:bestfit}
    \end{subfigure}%
    \hfill
    \begin{subfigure}[b]{0.50\textwidth}
        \centering
        \includegraphics[width=80mm]{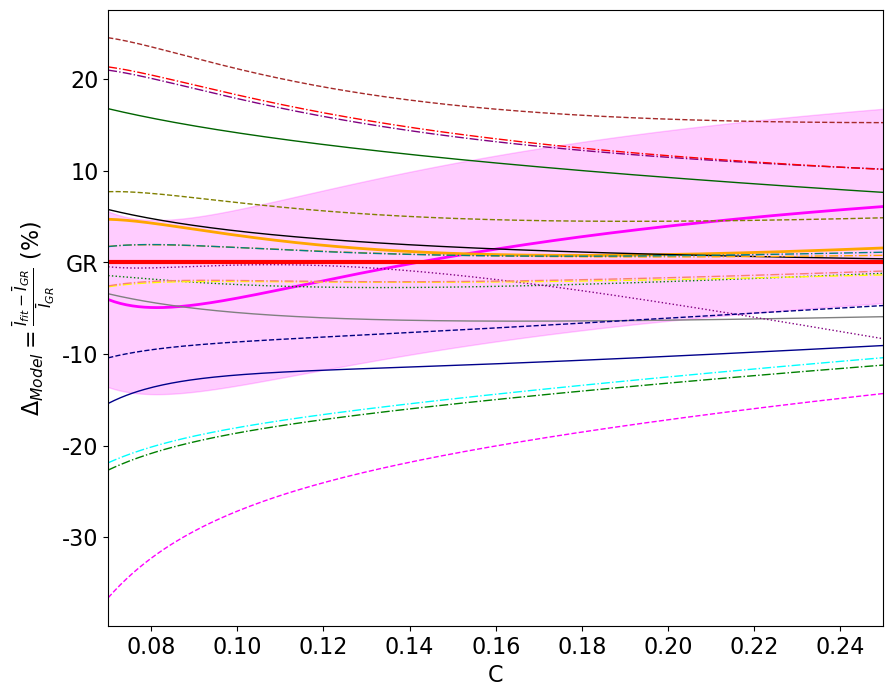}
        \caption{The percentage deviation from the GR.}
        \label{fig:devfit}
    \end{subfigure}%
    \caption{The best-fitting $\bar{I}-C$ relation for GR and all $f(Q)$ models with various parameters $\alpha$ and $\beta$. The percentage deviation of the best-fitting $\bar{I}-C$ relation for each model from the GR best-fitting relation in this calculation is also shown. The magenta and pink line corresponds to the $\bar{I}-C$ relation calculation for PSR J0737-3039A from \citet{Miao_2021_515_5071} with error estimation about $\sim 10\%$. The orange solid line represents the difference between our GR calculation and the results from \citet{Breu_2016_459_646}, which is approximately 4\%.}
    \label{fig:fitcombine}
\end{figure*}

By calculating the moment of inertia, mass, and radius of neutron stars, we successfully computed the $\bar{I}-C$ relation for the four extended $f(Q)$ gravity models. The results for each model and EoS are shown in Fig. \ref{fig:fitline}. All results show a similar overall trend to that reported by \citet{Breu_2016_459_646}, and are consistent with the pattern obtained by \citet{Suleiman:2024ztn} for nuclear-physics-based EoS sets, indicating consistent qualitative behavior of the $\bar{I}$–$C$ relation. In the bottom panels of each graph, we present the residuals from the fitting, defined as $\Delta = (\bar{I}_{\text{fit}} - \bar{I})/\bar{I}$, where all models exhibit a maximum residual of around 10\%. The best-fit for each model is plotted in Fig. \ref{fig:fitcombine} along with its deviation from GR. The calculation results demonstrate consistency with their respective $\bar{\omega}$ profiles and $I$ properties. When compared with the mass-radius relation, the $\bar{I}-C$ relation also shows consistent results. For smaller masses, as observed in the quadratic model or when the additional terms are negative in the logarithmic model, the deviations of $\bar{I}$ are positive. However, for larger masses, the deviations of $\bar{I}$ become negative. In contrast, the exponential model exhibits the opposite behavior. When the neutron star has a larger mass, the deviation of $\bar{I}$ is positive, whereas for smaller masses, the deviation is negative.

However, as emphasized by \citet{Suleiman:2024ztn}, the apparent universality of this relation is significantly influenced by the variability of the EoS. This is also reflected in the residual distribution $\Delta$ in Fig.~\ref{fig:fitline}, where each model exhibits distinct patterns: for instance, the exponential model appears more sensitive to the Skyrme EoS family, while the logarithmic model shows stronger deviations with RMF-based EoS. The quadratic model, on the other hand, tends to be more universal at low compactness, with residuals remaining within about $4\%$. Nevertheless, all models display a similar trend at high compactness, with residuals increasing up to about $10\%$. \citet{Suleiman:2024ztn} further demonstrated, through a comparison between nuclear-physics-based and agnostic (metamodel and Gaussian-process) EoS ensembles, that the $\bar{I}$–$C$ relation can exhibit up to $\sim20\%$ dispersion at high compactness when broader EoS families are considered, whereas the $\bar{I}$–$\Lambda$ relation remains substantially more universal. This dispersion originates from uncertainties in nuclear microphysics, such as the onset of hyperonic or quark degrees of freedom, which are not fully captured when only a limited set of EoS is employed. Consequently, if the EoS variability is insufficiently explored, any deviation observed in $\bar{I}$–$C$ could be misinterpreted as a signature of modified gravity, leading to a degeneracy between gravitational and dense-matter effects.

\begin{figure*}
    \centering
    \begin{subfigure}[b]{0.50\textwidth}
        \centering
        \includegraphics[width=80mm]{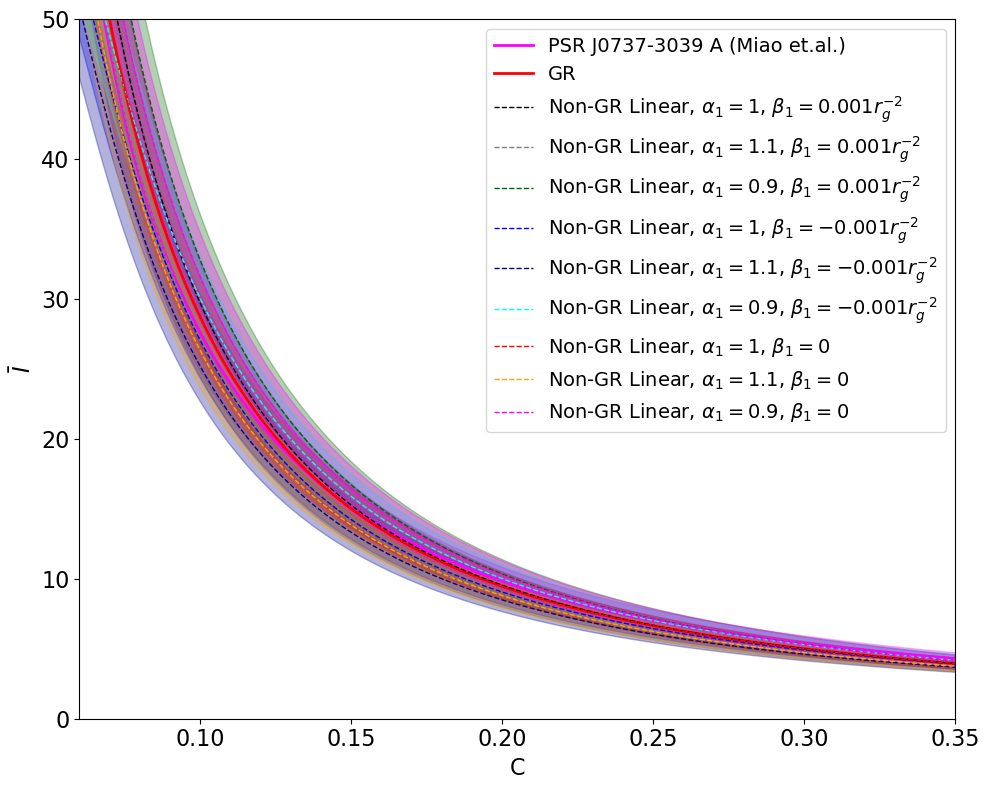}
        \caption{The best-fitting $\bar{I}-C$ relation of $f(Q)=\alpha_{1} Q+\beta_{1}$}
        \label{fig:bestfitlin}
    \end{subfigure}%
    \hfill
    \begin{subfigure}[b]{0.50\textwidth}
        \centering
        \includegraphics[width=80mm]{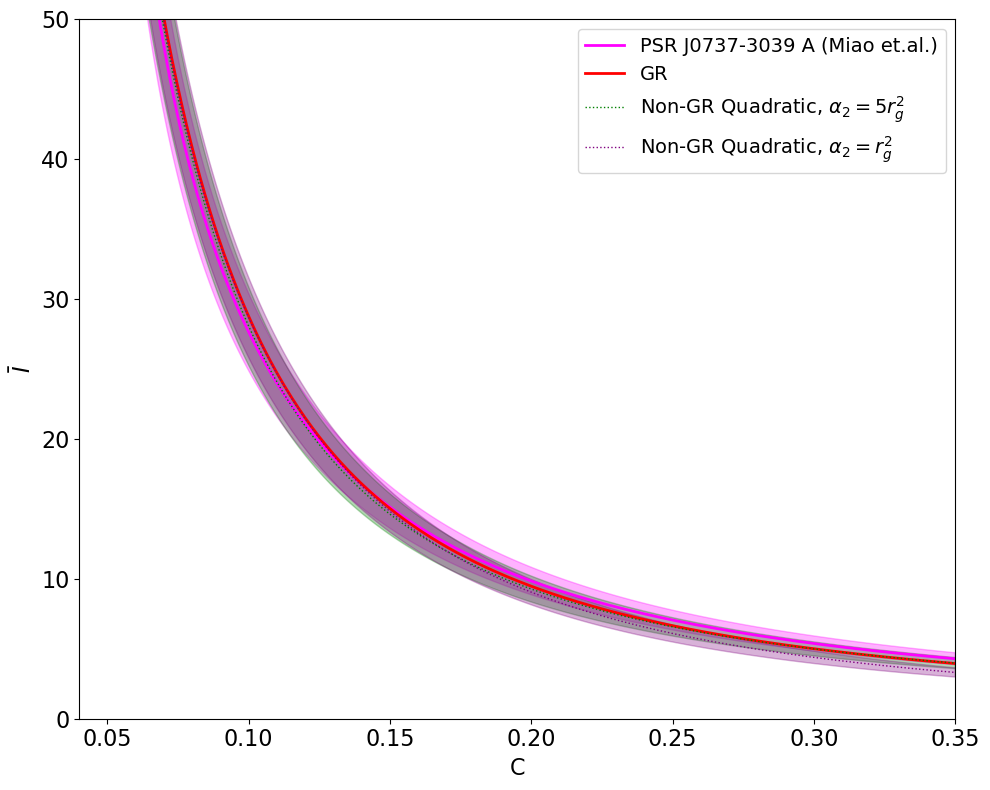}
        \caption{The best-fitting $\bar{I}-C$ relation of $f(Q)=Q+\alpha_{2} Q^2$}
        \label{fig:bestfitquad}
    \end{subfigure}%
    \vskip\baselineskip
    \begin{subfigure}[b]{0.50\textwidth}
        \centering
        \includegraphics[width=80mm]{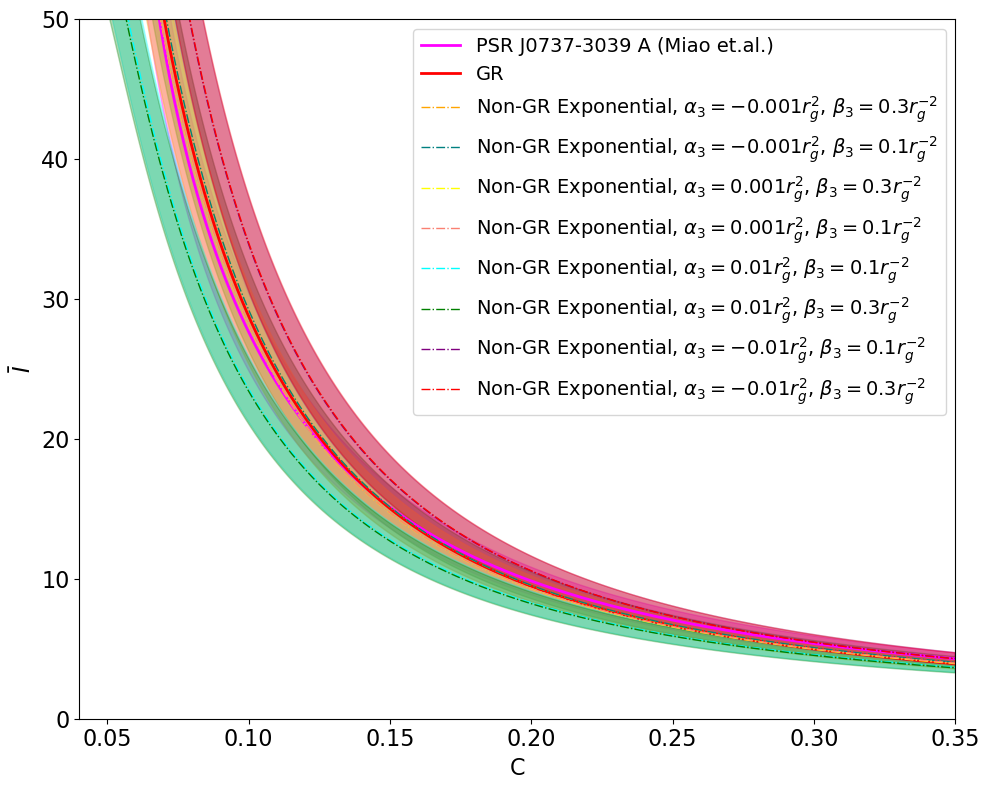}
        \caption{The best-fitting $\bar{I}-C$ relation of $f(Q) = Q + \alpha_{3} e^{\beta_{3} Q}$}
        \label{fig:bestfitexp}
    \end{subfigure}%
    \hfill
    \begin{subfigure}[b]{0.50\textwidth}
        \centering
        \includegraphics[width=80mm]{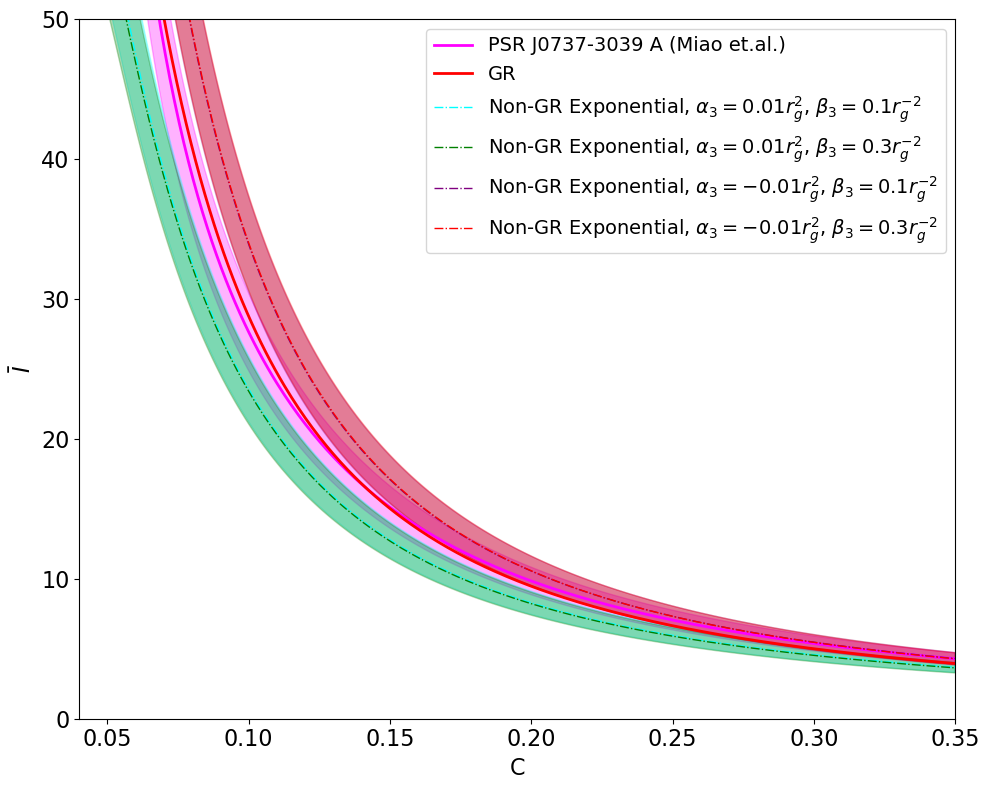}
        \caption{The best-fitting $\bar{I}-C$ relation of $f(Q) = Q + \alpha_{3} e^{\beta_{3} Q}$, which lies outside the PSR J0737–3039A band.}
        \label{fig:bestfitexpoutside}
    \end{subfigure}
    \vskip\baselineskip
    \begin{subfigure}[b]{0.50\textwidth}
        \centering
        \includegraphics[width=80mm]{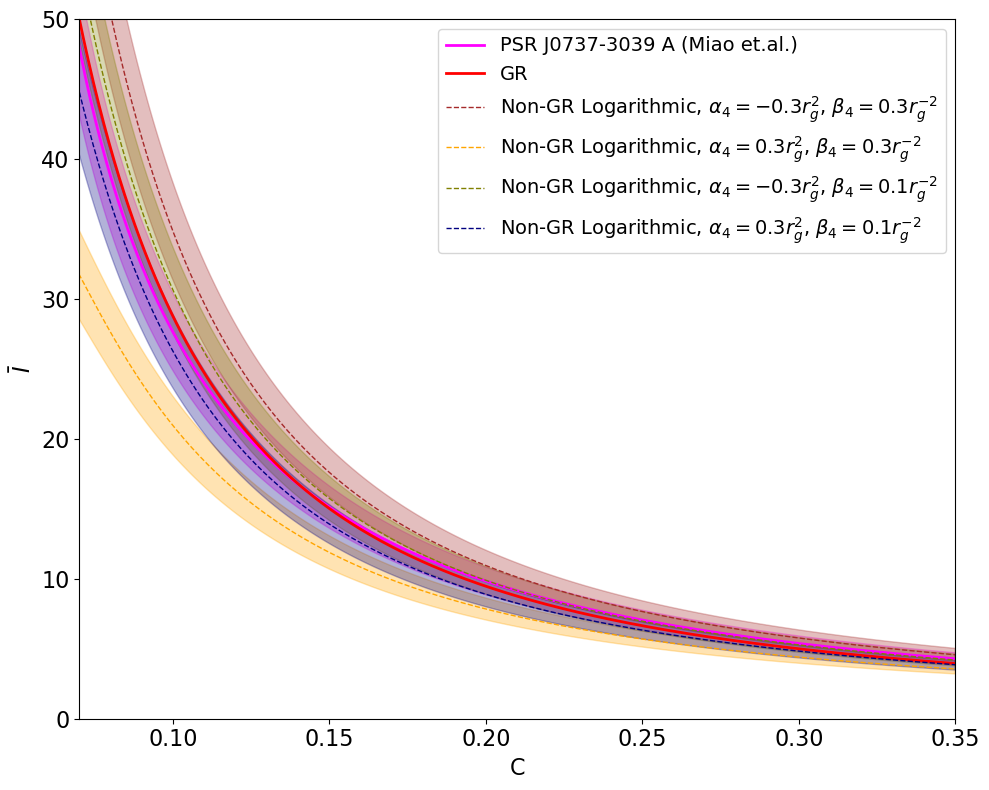}
        \caption{The best-fitting $\bar{I}-C$ relation of $f(Q) = Q - \alpha_{4} \ln(1-\beta_{4} Q)$}
        \label{fig:bestfitlog}
    \end{subfigure}%
    \hfill
    \begin{subfigure}[b]{0.50\textwidth}
        \centering
        \includegraphics[width=80mm]{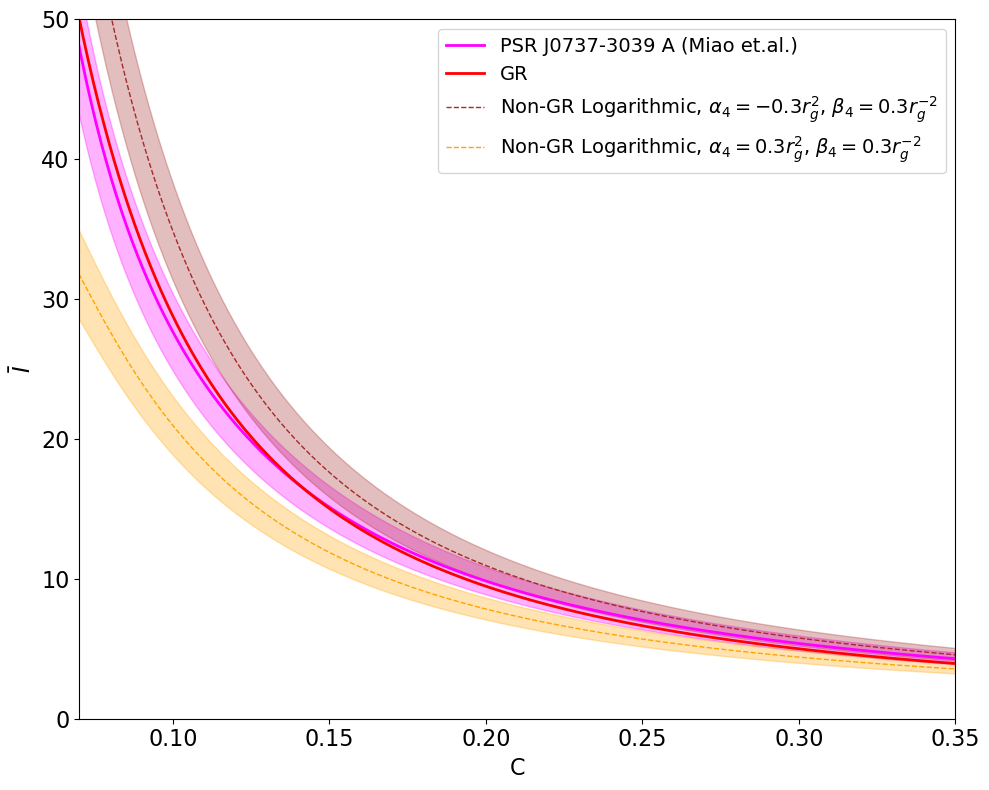}
        \caption{The best-fitting $\bar{I}$–$C$ relation of $f(Q) = Q - \alpha_{4} \ln(1-\beta_{4} Q)$, which lies outside the PSR J0737–3039A band.}
        \label{fig:bestfitlogoutside}
    \end{subfigure}
    \caption{The best-fitting $\bar{I}$--$C$ relations for all $f(Q)$ models with an estimated EoS variability of 4-10\%. The shaded regions represent the corresponding $\bar{I}$--$C$ bands, showing partial overlaps among models. These results highlight the sensitivity of the $\bar{I}$--$C$ relation to both the gravitational modification parameters and the EoS uncertainty.}
    \label{fig:bestfitline}
\end{figure*}

In our results, this degeneracy manifests clearly. The quadratic $f(Q)$ model, shown in Fig.~\ref{fig:bestfitquad}, produces only marginal deviations from GR, remaining within the $\lesssim10\%$ scatter range consistent with the EoS variability reported by \citet{Suleiman:2024ztn}. The linear model, shown in Fig.~\ref{fig:bestfitlin}, exhibits moderate deviations depending on the coupling parameter $\alpha$, but these still lie within the typical $\lesssim20\%$ EoS-induced variability band. On other hand, the logarithmic (Fig.~\ref{fig:bestfitlog}) and exponential (Fig.~\ref{fig:bestfitexp}) models display substantially larger departures from GR, exceeding the expected EoS-driven scatter and thus offering potential discriminators between the effects of modified gravity and dense-matter microphysics.
Notably, the exponential model exhibits an amplified response even for small parameters. For instance, with $\alpha = 0.01r_g^{-2}$, the deviation reaches approximately $20\%$ at $C = 0.08$ relative to GR, while it gradually diminishes at higher compactness. This strong sensitivity implies that even minor variations in the parameter can lead to significant departures from GR, highlighting the model’s non-linear nature. Furthermore, such behavior may indicate potential instability of neutron stars in the exponential model due to rotational effects, as suggested by the characteristics of $\bar{\omega}$, the moment of inertia, and this relation. Although previous studies have shown that the exponential model can support neutron stars up to $2.8,M_\odot$ for $\alpha = r_g^{-2}$ and $\beta = 0.2r_g^{2}$, our results demonstrate that the $\bar{I}$–$C$ relation remains consistent only for much smaller parameters, such as $\alpha = 0.01r_g^{-2}$.
This pronounced parameter sensitivity suggests that precise $\bar{I}$–$C$ measurements could place stringent constraints on the exponential model and may point to possible instabilities in rapidly rotating configurations, as reflected in the $\bar{\omega}$ and $I$ profiles. In light of these results and following the conclusions of \citet{Suleiman:2024ztn}, we emphasize that quasi-universal relations involving the compactness such as $\bar{I}$–$C$ are susceptible to EoS variability, whereas those involving tidal deformability (e.g., $\bar{I}$–$\Lambda$) remain more robust. Therefore, any attempt to use the $\bar{I}$–$C$ relation as a probe of gravity must properly account for EoS uncertainty to avoid degeneracy. For the quadratic and linear $f(Q)$ models, the deviations are largely dominated by EoS variability, while in the logarithmic and exponential models, the larger departures exceed the quasi-universal scatter. This indicates that future multimessenger observations, particularly those capable of measuring the moment of inertia with $\sim5\%$ precision, such as PSR~J0737–3039A or third-generation gravitational-wave detectors could effectively break this degeneracy.

The fitting coefficients derived from the best-fit analysis are summarized in Table~\ref{tab:fitting_coefficients}. From the table, we can observe how the variations in the parameters $\alpha_i$ and $\beta_i$ translate into positive or negative deviations relative to the GR fitting coefficients, consistent with the behavior illustrated in Fig.~\ref{fig:fitcombine}. In that figure, the calculated $\bar{I}$–$C$ relations for the four $f(Q)$ models are compared with the observationally constrained $\bar{I}$–$C$ relation of PSR J0737–3039A, considered as a neutron star~(\cite{Miao_2021_515_5071}), with an estimated uncertainty of approximately $10\%$. It is found that the linear and quadratic models remain fully consistent within the $\bar{I}$–$C$ band of PSR J0737–3039A across the entire compactness range, indicating that moderate geometric modifications do not significantly alter the universal trend predicted by GR. By comparison, for the logarithmic and exponential models, certain parameter combinations produce noticeable deviations from the observational band, particularly at low compactness. As shown in Fig.~\ref{fig:bestfitexpoutside} and Fig.~\ref{fig:bestfitlogoutside}, the exponential model with  $\alpha_3=\pm 0.01 r_g^2$ values and the logarithmic model with $\alpha_4 = \pm 0.3~r_g^2$ and $\beta_4 = 0.3~r_g^{-2}$ exhibit $\bar{I}$–$C$ curves that extend beyond the GR-predicted region. Although these deviations place the models partially outside the empirical band, they reveal the high sensitivity of such modified gravity formulations to stellar structure at lower densities, where nonlinear geometric corrections become more pronounced. Rather than indicating inconsistency, these departures may point to distinctive features inherent to the logarithmic and exponential forms of $f(Q)$ gravity, suggesting that such models could capture subtle deviations from GR while remaining compatible with observations at higher compactness.

We have also explored the EoS with different particle contents, including pions in PS and quarks and hyperons in PCL2 and SQM, which are EoS models for strange stars. In both PCL2 and PS, the calculation results maintain universality. However, the residuals increase by up to 20\% at higher compactness due to the dispersion of data points. This may occur, one possible reason being that this EoS is more suitable for anisotropic NS (\cite{Sawyer:1972cq, Canuto:1974ft, Das:2022ell}). For SQM, the behavior is similar to what is observed in other modified gravity models with changes to the geometric components, such as $f(R)$ modifications. At low compactness, strange stars deviate from the universal relation, although their $I$ values converge toward the neutron star relation at high compactness, indicating the need for further investigation. Therefore, we present the results here without fully considering the characteristics of strange stars at this stage. These results are certainly interesting for future investigations, especially those involving calculations in $f(Q)$ theory using SQM or EoS of quark stars, which are excluded from the current discussion, as stated previously.


\section{Conclusion}\label{Sec4}
In this paper, we calculated NS properties within the framework of $f(Q)$ gravity. First, we derived the modified TOV equations using slowly rotating metric and determined the initial values and integration limits for these equations. Then, we obtained numerical solutions using the shooting method with realistic EoS and computed the profiles of $\bar{\omega}$ and the moment of inertia for four extended $f(Q)$ models: linear, quadratic, logarithmic, and exponential. Based on the results, we tested their consistency with the mass-radius relation and profiles obtained in our previous work, finding that the solutions align well with the angular velocity $\bar{\omega}$ and the corresponding moment of inertia. Finally, we calculated the universal relation between the moment of inertia and the compactness of the star.

Non-metricity, as the main geometric component in this study, influences the distribution of the geometric profiles $A(r)$ and $B(r)$, leading to deviations in the properties of the star. This also affects the angular velocity and the moment of inertia of NS in the slowly rotating case. In $f(Q)$ models, deviations in the angular velocity are observed at the center of the star, with larger deviations corresponding to higher values of the $f(Q)$ parameter. A similar trend is observed in the moment of inertia, where the deviations are even more pronounced than those in the maximum mass. This makes the moment of inertia a potentially good constraint for gravitational theories, offering the possibility to probe or even rule them out. We further explored the universal $\bar{I}$–$C$ relation, finding that using $I/M^3$ as a function of compactness demonstrates an approximately universal trend within the $f(Q)$ theory, with residuals around 10\%. While this level of universality is weaker than the subpercent accuracy achieved in the $I$–Love–$Q$ relations (\cite{Yagi:2013awa, Yagi:2013bca}), it still indicates that the $\bar{I}$–$C$ relation in $f(Q)$ gravity preserves a similar qualitative behavior to GR. The relatively small difference of about 4\% between our GR calculation and \citet{Breu_2016_459_646} confirms the numerical consistency of our results, even though the scatter across different EoS remains larger in the $f(Q)$ framework. This $\sim4\%$ deviation from \citet{Breu_2016_459_646} result is reasonable, as our analysis includes a significantly larger and more diverse set of EoS, which naturally broadens the fitting range. A similar level of deviation has also been reported by \citet{Suleiman:2024ztn}, who found comparable differences when extending the $\bar{I}$–$C$ analysis to a wider EoS set. In the slowly rotating case, the rotational effects in each model exhibit consistent behavior with the obtained values of $I$ and correlate well with the mass-radius relation. However, the exponential model is notable, as its rotational effects are relatively more pronounced, showing some deviations when $\alpha$ exceeds $0.01 r_g^2$. This presents a limitation for the exponential model, especially if the observed $\bar{I}-C$ relation shows minimal deviation from GR. In contrast, models that show relatively large deviations in maximum mass, such as the logarithmic model, demonstrate more stable and consistent deviations as the parameters change, as illustrated in Fig. \ref{fig:devfit}.

In this paper, we also try to constrain the $\bar{I}-C$ relation using observational data from binary systems containing pulsars, such as $\bar{I}-C$ relation for the PSR J0737-3039A system (\cite{Miao_2021_515_5071}). By using gravitational wave data from LIGO/Virgo (GW170817 and GW190425), which constrain tidal deformability ($\Lambda$), and mass-radius measurements from NICER (PSR J0030+0415 and PSR J0740+6620), it is possible to estimate the moment of inertia of PSR J0737-3039A. Bayesian analysis suggests that the moment of inertia of this pulsar is approximately $\sim 1.30 \times 10^{45} \text{g cm}^2$, depending on the employed hadronic EoS. Furthermore, these measurements allow the study of possible phase transitions in the core of pulsars, such as transitions from hadronic to quark matter. If PSR J0737-3039A is a quark star, its moment of inertia is expected to be larger, around $\sim 1.55 \times 10^{45} , \text{g cm}^2$. This suggests that accurate measurements of $I$ could help distinguish between hadronic neutron stars, hybrid stars, and quark stars. The $\bar{I}-C$ results obtained are consistent with the behavior observed in the calculation of the $\bar{I}-C$ relation for PSR J0737-3039A, where the moment of inertia of a quark star deviate larger than that of a (hybrid-)neutron star in low compactness, in agreement with references in alternative gravity theories such as $f(R)$.

With more precise measurements of the moment of inertia in the future, this universal $\bar{I}$–$C$ relation can be used to test the consistency of observational data with GR or alternative gravity theories, while simultaneously constraining the EoS of dense matter. As shown in Fig.~\ref{fig:bestfit}, the $\bar{I}$–$C$ relation for PSR~J0737–3039A provides a useful benchmark for constraining the parameters $\alpha_i$ and $\beta_i$ in $f(Q)$ gravity. In our analysis, the linear and quadratic models exhibit $\bar{I}$–$C$ curves that remain entirely within the observational band of PSR~J0737–3039A, suggesting that mild geometric modifications preserve the universality predicted by GR. On other hand, the logarithmic and exponential models display moderate deviations from the GR band at low compactness, as illustrated in Fig.~\ref{fig:bestfitexpoutside} and Fig.~\ref{fig:bestfitlogoutside}. Such deviations, however, should not be interpreted as inconsistencies. Previous studies~(\cite{Suleiman:2024ztn}) have shown that the apparent mismatch between theoretical $\bar{I}$–$C$ relations and observational estimates can also arise from residual EoS uncertainties, especially at low compactness where the EoS dependence becomes more pronounced. In this sense, the deviations observed here may reflect a combined effect of EoS variability and geometric nonlinearities inherent to the logarithmic and exponential $f(Q)$ models. Therefore, rather than contradicting the observational band, these results emphasize the importance of disentangling the influence of the EoS from that of the underlying gravitational theory when interpreting universal relations.

These results suggest that $f(Q)$ gravity remains a viable and testable framework in the strong-field regime. The obtained $\bar{I}$–$C$ relation demonstrates that modified geometric effects can lead to observable deviations while still remaining broadly consistent with the current empirical constraints. With tighter observational limits in the near future, such relations could place stringent bounds on the model parameters or rule out specific functional forms of $f(Q)$. For instance, continued timing observations of the Double Pulsar, PSR~J0737–3039A/B, are expected to yield a direct measurement of the moment of inertia of pulsar~A with an accuracy of about 11\% by 2030~(\cite{Kramer:2021jcw}). The increased sensitivity of upcoming radio facilities such as the Square Kilometre Array (SKA) will also be important, as they are expected to discover a larger population of double neutron star systems. A larger sample of such binaries would provide multiple data points to test the universality of the $I$--$C$ relation across a wider range of masses, offering a more robust method for testing gravitational theories.

Moreover, the universality of this relation is known to extend beyond the slowly rotating case, remaining valid for rapidly and even extremely rotating configurations in other gravity models~(\cite{Astashenok:2020cqq, Doneva:2016xmf}). Extending such an analysis within the $f(Q)$ framework would be an important next step, though more complex due to non-metricity corrections appearing at $\mathcal{O}(\epsilon^2)$ and higher orders.  In addition, the present results motivate further investigation using more accurate and EoS-insensitive quasi-universal relations, such as the $I$–$\Lambda$, $I$–$Q$, and $I$–Love–$C$ correlations~(\cite{Yagi:2013awa, Yagi:2013bca, Maselli:2013mva, Yagi:2016bkt, Jiang:2020uvb}), which can provide tighter and complementary constraints than the $\bar{I}$–$C$ relation. In particular, \citet{Suleiman:2024ztn} have shown that, while the $\bar{I}$–$C$ relation exhibits a mild dependence on the choice of EoS in GR, the $I$–$\Lambda$ and $C$–$\Lambda$ relations display significantly reduced variability, typically below the percent level, even when agnostic EoS datasets are considered. This indicates that the level of universality itself, rather than the best-fit curve alone, can serve as a useful diagnostic for probing possible deviations from GR. Consequently, future extensions of the present work could explore whether such reduced variability persists within the $f(Q)$ framework, where geometric corrections might introduce measurable departures from the GR trend. Furthermore, a recently proposed EoS-insensitive relation by~\cite{Papigkiotis:2025cjy} links the polar-to-equatorial radius ratio $\mathcal{R}=R_{\mathrm{pole}}/R_{\mathrm{eq}}$ to the compactness $C=M/R_{\mathrm{eq}}$ and reduced spin $\sigma=\Omega^3 R_{\mathrm{eq}}^3/GM$. Using an artificial neural network trained on 70 different EoS, this relation achieves an accuracy better than $0.25\%$, offering a promising complementary approach for testing gravity theories beyond the $\bar{I}$–$C$ relation framework.

\section*{Acknowledgments}
The authors thank R. Saito and N. Yoshioka for insightful discussions. MAA also thanks M. D. Danarianto for valuable guidance. BM acknowledges support from IUCAA, Pune (India), during an academic visit in which this work was carried out. The authors also sincerely thank the anonymous reviewers for their constructive comments and suggestions, which have greatly helped improve the clarity and quality of this article.

\section*{Data Availability}
There are no associated data with this article. No new data were generated or analysed in support of this research.

\bibliographystyle{mnras.bst}
\bibliography{References}

\appendix
\section{Energy-Momentum Conservation Constraints in $f(Q)$ Gravity}

One of the concerns in covariant $f(Q)$ gravity under the spherically symmetric metric is the constraint from energy-momentum conservation, especially for non-linear forms of $f(Q)$. Following Refs. \cite{Zhao_2022_82_303} and \cite{De_2023_40_115007}, these constraints originate from the left-hand side of Eq.~\eqref{eq8}, which should vanish identically when combined with the covariant derivative of the energy-momentum tensor $\mathring{\nabla}_\mu \mathcal{T}^{\mu\nu} = 0 $. Let us re-evaluate both sides of Eq.~\eqref{eq8}, employing the slowly rotating metric as defined in Eq.~\eqref{eq10}. On the right-hand side, we obtain:
\begin{align}
\mathring{\nabla}_\mu \mathcal{T}^{\mu\nu}
&= \partial_\mu T^{\mu\nu} + \Gamma^\mu_{\mu\lambda} T^{\lambda\nu} + \Gamma^\nu_{\mu\lambda} T^{\mu\lambda}, \nonumber\\
\mathring{\nabla}_\mu \mathcal{T}^{\mu r}
&= \partial_r T^{rr} + \left( \Gamma^t_{tr} + \Gamma^r_{rr} + \Gamma^\theta_{\theta r} + \Gamma^\phi_{\phi r} \right) T^{rr} \nonumber\\
&\quad + \Gamma^r_{tt} T^{tt} + \Gamma^r_{rr} T^{rr} + \Gamma^r_{\theta\theta} T^{\theta\theta} + \Gamma^r_{\phi\phi} T^{\phi\phi} + 2\Gamma^r_{t\phi} T^{t\phi}, \nonumber\\
&= e^{-B} \left[ p' + \frac{(p + \rho c^2)}{2} A' + \mathcal{O}(\epsilon^2) \right], \nonumber\\
&= e^{-B} \left[ p' + \frac{(p + \rho c^2)}{2} A'\right].
\end{align}
Assuming $\mathring{\nabla}_\mu \mathcal{T}^{\mu\nu} = 0$, we find that the continuity equation remains unchanged. Next, we explicitly compute the full left-hand side of Eq.~\eqref{eq8} by rewriting it such that the gravitational sector is denoted by $E_{\mu\nu}$ on the left-hand side, while the matter sector is represented by $\mathcal{T}_{\mu\nu}$ on the right-hand side.
 \begin{align}
    E_{\mu\nu} \equiv f_{Q}\mathring{G}_{\mu\nu}+\frac{1}{2}g_{\mu\nu}(Qf_{Q}-f)+2f_{QQ}P^{\lambda}_{~~\mu\nu}\mathring{\nabla}_{\lambda}Q &= \kappa \mathcal{T}_{\mu \nu}~.
\end{align}
Following \cite{Zhao_2022_82_303}, the covariant derivative of the left-hand side leads to a constraint given by:
\begin{equation}\label{apA}
    \frac{(e^B - 1)(4 + rA' + rB') + 2rB'}{2r^2} f_Q' + \frac{(e^B - 1)}{r} f_Q'' = \mathring{\nabla}_\mu \mathcal{T}^{\mu\nu}~.
\end{equation}
where $ f_Q' = f_{QQ} \frac{dQ}{dr} = f_{QQ} Q'$ and $ f_Q'' = f_{QQQ} \left( \frac{dQ}{dr} \right)^2 + f_{QQ} \frac{d^2Q}{dr^2}=f_{QQQ} Q'^2 + f_{QQ} Q''$. So eq.~(\ref{apA}) becomes
\begin{align}
&\left[ \frac{(e^B - 1)(4 + rA' + rB') + 2rB'}{2r^2} \right] f_{QQ} Q'\nonumber\\ 
&+ \frac{(e^B - 1)}{r} \left( f_{QQQ} Q'^2 + f_{QQ} Q'' \right)= e^{-B} \left[ p' + \frac{(p + \rho c^2)}{2} A'\right].
\end{align}
Switching to dimensionless form using Eq. \eqref{dimensionless} and Eq. \eqref{constant}, we get
{\fontsize{9pt}{9pt}
\begin{align}
&\frac{f_{QQ}}{r_g^5} \left[ \frac{(e^B - 1)\hat{r}\hat{A}' - (e^B + 1)\hat{r}\hat{B}' + 4(e^B - 1)}{2 \hat{r}^2} \hat{Q}' 
+ \frac{e^B - 1}{\hat{r}} \hat{Q}'' \right]\nonumber\\
&\quad\quad\quad\quad\quad+\frac{f_{QQQ}}{r_g^5} \left[\frac{(e^B - 1)}{\hat{r}} \hat{Q}'^2\right] 
= \frac{p_g}{r_g} e^{-B} \left[ \hat{p}' + \frac{(\hat{p} + \hat{\rho})}{2} \hat{A}' \right],\nonumber\\[10pt]
&\frac{f_{QQ}}{p_g r_g^4} \left[ e^B \left(\frac{(e^B - 1)\hat{r}\hat{A}' - (e^B + 1)\hat{r}\hat{B}' + 4(e^B - 1)}{2 \hat{r}^2} \hat{Q}' 
+ \frac{e^B - 1}{\hat{r}} \hat{Q}'' \right)\right]\nonumber\\
&\quad\quad\quad\quad\quad+ \frac{f_{QQQ}}{p_g r_g^4} \left[e^B \frac{(e^B - 1)}{\hat{r}} \hat{Q}'^2\right] 
= \hat{p}' + \frac{(\hat{p} + \hat{\rho})}{2} \hat{A}'.\nonumber
\end{align}}
Let us define
\begin{align}
    \Phi_r &\equiv e^B \left[ \frac{(e^B - 1)\hat{r}\hat{A}' - (e^B + 1)\hat{r}\hat{B}' + 4(e^B - 1)}{2 \hat{r}^2} \hat{Q}' + \frac{e^B - 1}{\hat{r}} \hat{Q}'' \right],\\[5pt]
    \Psi_r &\equiv e^B \frac{(e^B - 1)}{\hat{r}} \hat{Q}'^2.
\end{align}
Therefore
\begin{align}
    \frac{\left[f_{QQ} \Phi_r + f_{QQQ} \Psi_r\right]}{p_g r_g^4} &= \hat{p}' + \frac{(\hat{p} + \hat{\rho})}{2} \hat{A}',\nonumber\\
    \frac{\left[f_{QQ} \Phi_r + f_{QQQ} \Psi_r\right]}{M_\odot c^2 r_g} &= \hat{p}' + \frac{(\hat{p} + \hat{\rho})}{2} \hat{A}'.
\end{align}
Since the left-hand side is found to be numerically small, we have:
\begin{align}
   0 \approx \hat{p}' + \frac{(\hat{p} + \hat{\rho})}{2} \hat{A}' \Rightarrow \hat{p}' \approx -\frac{(\hat{p} + \hat{\rho})}{2} \hat{A}'. 
\end{align}
From the above calculations, it is clear that the constraints in gravity $f(Q)$ can be effectively absorbed or suppressed when expressed in terms of dimensionless physical variables in the context of neutron stars. This makes the gravitational covariance formulation of $f(Q)$ more flexible, especially for non-linear models.

\begin{figure*}
    \centering
    \begin{subfigure}[b]{0.32\textwidth}
        \centering
        \includegraphics[width=\linewidth]{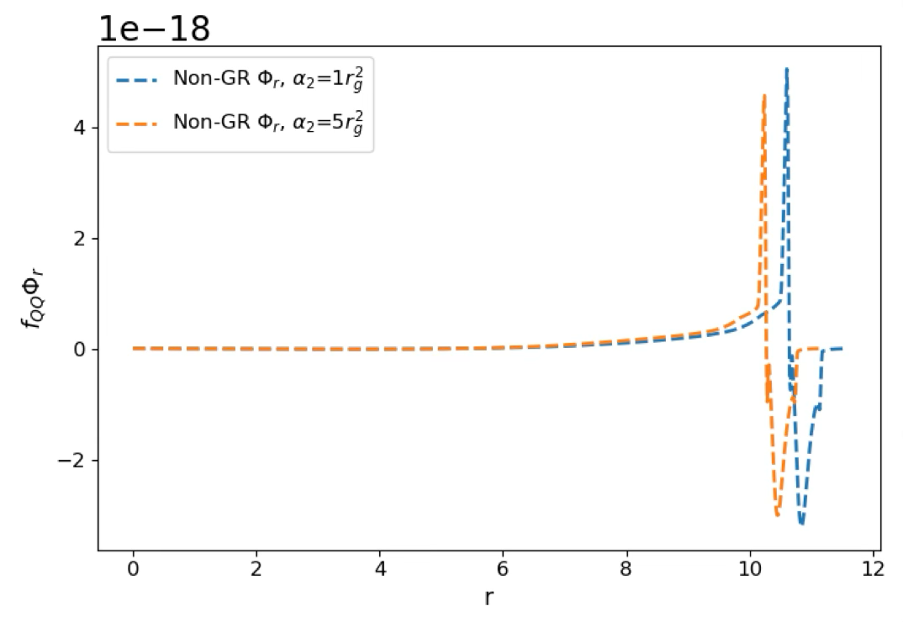}
        \caption{$f_{QQ}\Phi_r$ of $\alpha_2 Q^2$ model}
    \end{subfigure}%
    \hfill
    \begin{subfigure}[b]{0.32\textwidth}
        \centering
        \includegraphics[width=\linewidth]{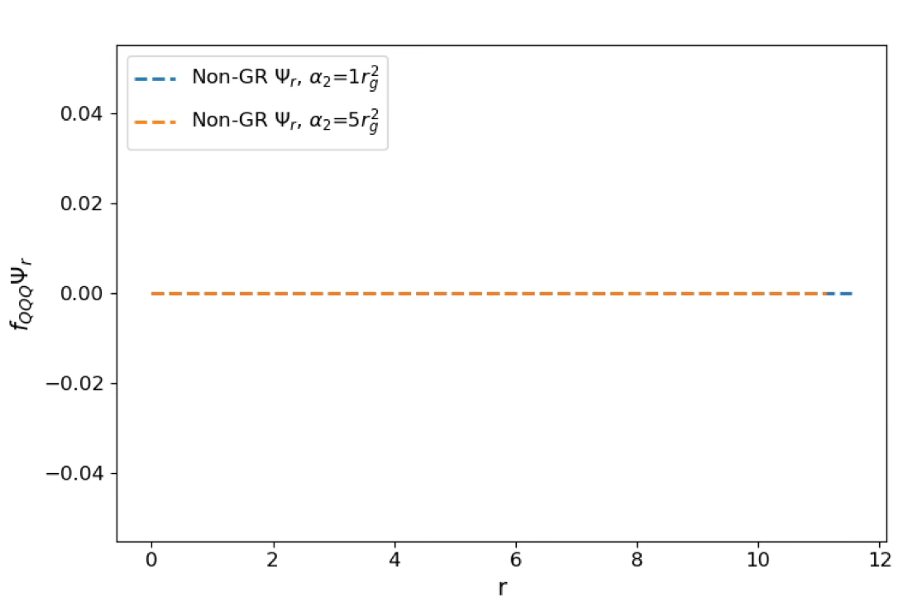}
        \caption{$f_{QQQ}\Psi_r$ of $\alpha_2 Q^2$ model}
    \end{subfigure}%
    \hfill
    \begin{subfigure}[b]{0.32\textwidth}
        \centering
        \includegraphics[width=\linewidth]{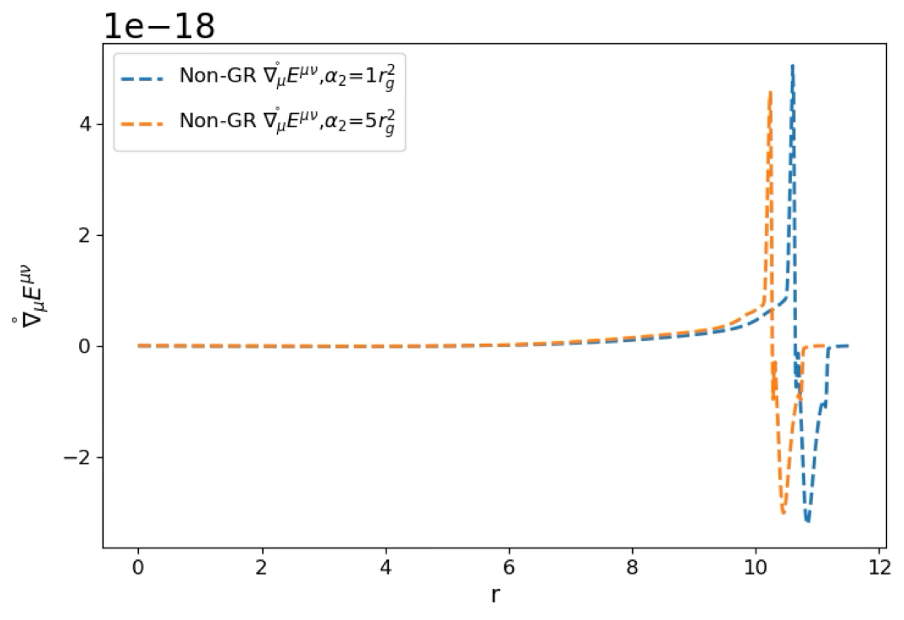}
        \caption{$\mathring{\nabla}_\mu E^{\mu\nu}$ of $\alpha_2 Q^2$ model}
    \end{subfigure}
    \hfill
    \begin{subfigure}[b]{0.32\textwidth}
        \centering
        \includegraphics[width=\linewidth]{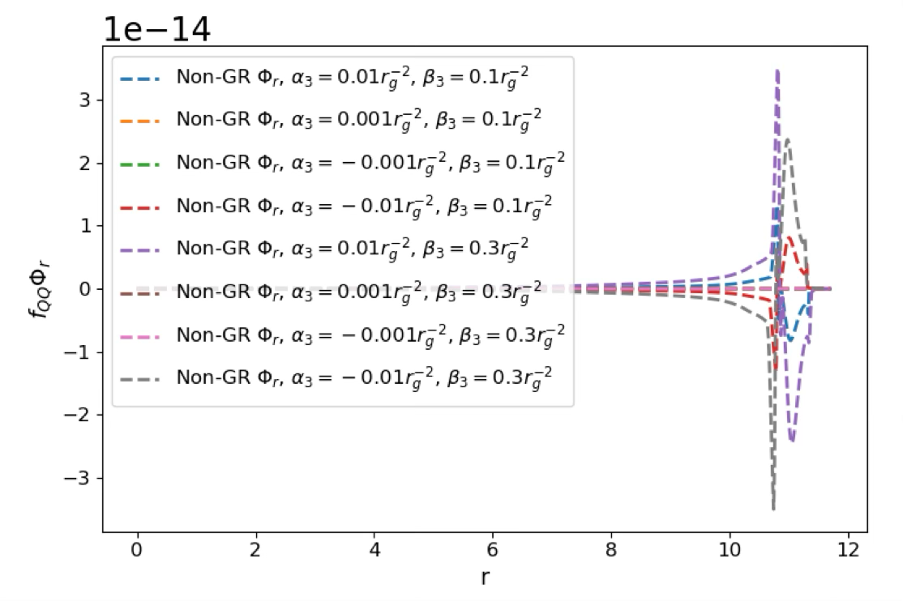}
        \caption{$f_{QQ}\Phi_r$ of $\alpha_3 e^{\beta_3 Q}$ model}
    \end{subfigure}%
    \hfill
    \begin{subfigure}[b]{0.32\textwidth}
        \centering
        \includegraphics[width=\linewidth]{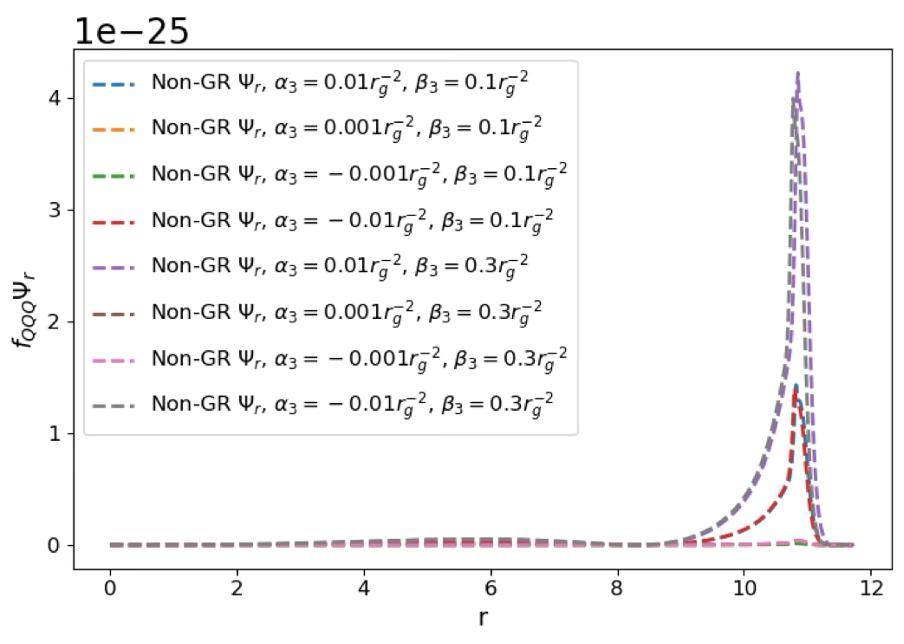}
        \caption{$f_{QQQ}\Psi_r$ of $\alpha_3 e^{\beta_3 Q}$ model}
    \end{subfigure}%
    \hfill
    \begin{subfigure}[b]{0.32\textwidth}
        \centering
        \includegraphics[width=\linewidth]{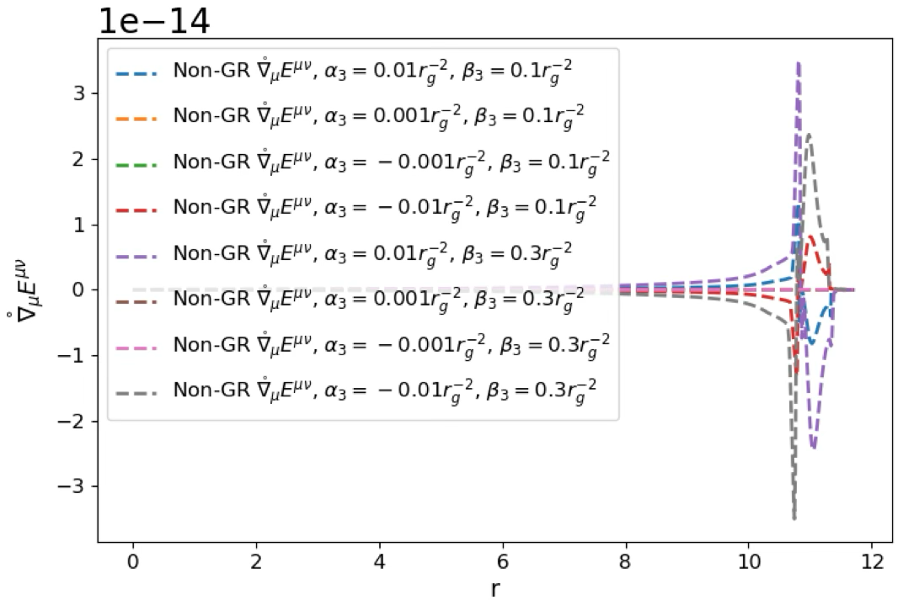}
        \caption{$\mathring{\nabla}_\mu E^{\mu\nu}$ of $\alpha_3 e^{\beta_3 Q}$ model}
    \end{subfigure}
    \hfill
    \begin{subfigure}[b]{0.32\textwidth}
        \centering
        \includegraphics[width=\linewidth]{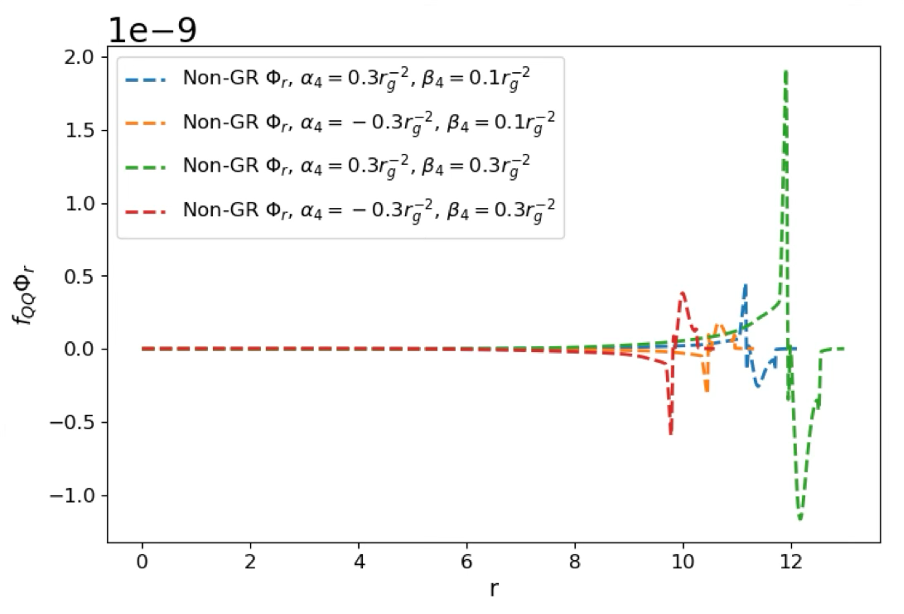}
        \caption{$f_{QQ}\Phi_r$ of $\alpha_4 \log (1-\beta_4 Q)$ model}
    \end{subfigure}%
    \hfill
    \begin{subfigure}[b]{0.32\textwidth}
        \centering
        \includegraphics[width=\linewidth]{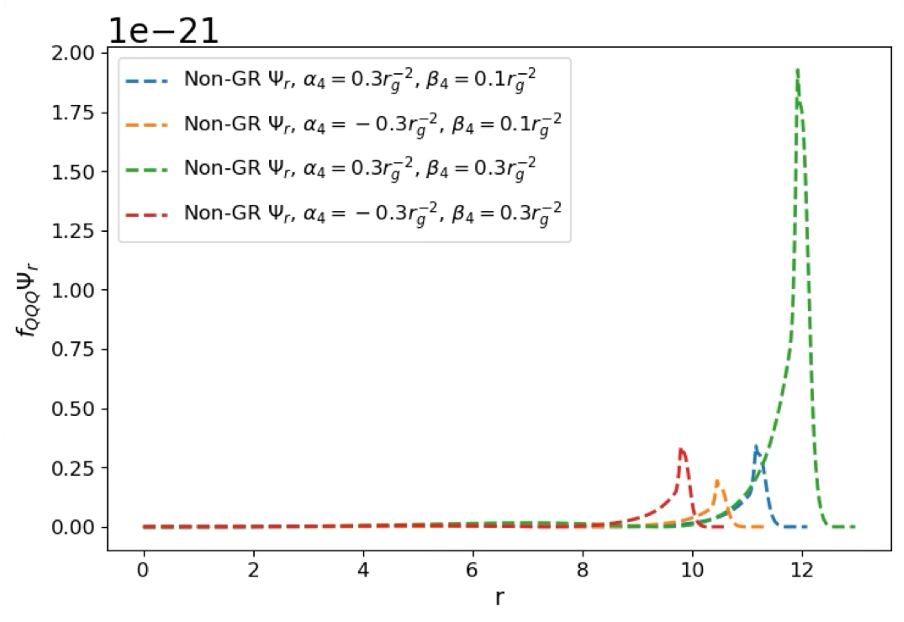}
        \caption{$f_{QQQ}\Psi_r$ of $\alpha_4 \log (1-\beta_4 Q)$ model}
    \end{subfigure}%
    \hfill
    \begin{subfigure}[b]{0.32\textwidth}
        \centering
        \includegraphics[width=\linewidth]{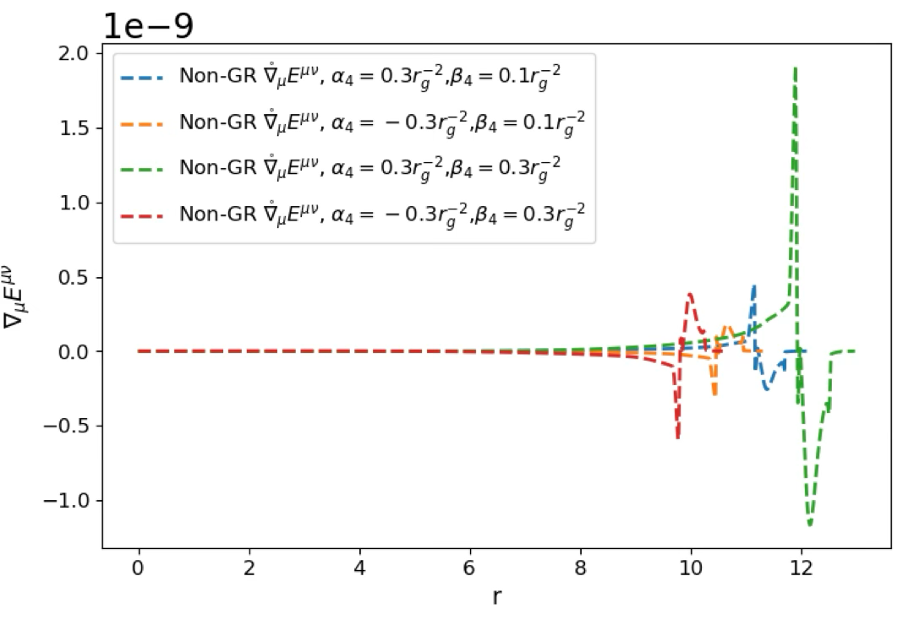}
        \caption{$\mathring{\nabla}_\mu E^{\mu\nu}$ of $\alpha_4 \log (1-\beta_4 Q)$ model}
    \end{subfigure}
    \caption{We present the profiles of $f_{QQ}\Phi_r$, $f_{QQQ}\Psi_r$, and $\mathring{\nabla}_\mu E^{\mu\nu}$ for each gravity model using the SLy equation of state, with a central density set at $\rho_c = 1 \times 10^{15}$ g/cm$^3$. It is important to note that the quadratic model does not include a $f_{QQQ}$ term. The plots indicate that $\mathring{\nabla}_\mu E^{\mu\nu}$ remains very small across all models, with the logarithmic model  reaching the highest order $\mathcal{O}(10^{-9})$. These results are consistent with the analytical calculations, which show that $\mathring{\nabla}_\mu E^{\mu\nu}$ vanishes at both the stellar core and surface due to the absence of nonmetricity.}
    \label{fig:emu}
\end{figure*}

\begin{table*}
\caption{\label{tab:constraint} Lower (LB) and upper bounds (UB) of the constraint deviation $\mathring{\nabla}_\mu E^{\mu\nu}$ for the 53 EoS considered.}
\renewcommand{\arraystretch}{1.5}
\begin{tabular}{ccc}
\hline\hline
$f(Q)$ Model           & $\mathring{\nabla}_\mu E^{\mu\nu}_{LB}$        &$\mathring{\nabla}_\mu E^{\mu\nu}_{UB}$                      \\ \hline
$Q + \alpha_{2}Q^2$         & $-6.639\times 10^{-17}$                 & $8.335\times 10^{-17}$                     \\
$Q + \alpha_{3}~e^{\beta_{3} Q}$ & $-6.529\times10^{-13}$                 & $6.49\times10^{-13}$                 \\
$Q - \alpha_{4} \ln(1-\beta_{4} Q)$& $-1.949\times 10^{-8}$                 & $1.69\times 10^{-8}$            \\
\hline\hline
\end{tabular}
\end{table*}
We have also numerically evaluated this constraint, and the results are presented in Fig.~\ref{fig:emu}. We describe the behaviors of $f_{QQ} \Phi_r$, $f_{QQQ} \Psi_r$, and $\mathring{\nabla}_\mu E^{\mu\nu}$ for each model. The numerical plots show that at the stellar center and surface, the constraints vanish due to the absence of non-metricity. However, between the core and the surface, the constraints are non-zero but remain extremely small ranging from $\mathcal{O}(10^{-18})$ in the quadratic model to $\mathcal{O}(10^{-9})$ in the logarithmic model.  This is consistent with the previous analytical results. We have also performed calculations using the 53 EoS (Table\ref{tab:eos_details}) considered in this work. Among all cases, the largest deviations in the constraints for the three models are summarized in Table~\ref{tab:constraint}, with values ranging from $\mathcal{O}(10^{-17})$ to $\mathcal{O}(10^{-8})$. Given these results, it is reasonable to assume $\mathring{\nabla}_\mu \mathcal{T}^{\mu\nu} \approx 0$, since the constraint is only negligible. Furthermore, in calculations of the neutron star structure, especially in the continuity equation, the part of matter governed by the EoS dominates, which allows us to neglect this very small constraint term and use the standard form of the continuity equation in Eq.~\eqref{toveq}.

\section{Fitting Coefficients from the best-fit of each model}\label{table}
\begin{table*}
\caption{\label{tab:fitting_coefficients}Fitting coefficients for various $f(Q)$ models.}
\renewcommand{\arraystretch}{1.6}
\begin{tabular}{c|cc|cccc}
\hline\hline
\multicolumn{3}{c|}{Model Parameters} & \multicolumn{4}{c}{Fitting Coefficients} \\ \hline
$f(Q)$ Model & $\alpha_{i}$ & $\beta_{i}$ & $a_1$ & $a_2$ & $a_3$ & $a_4$ \\ \hline
$\textbf{GR}$ & - & - & 0.603489&  0.291524& $-7.004818\times$ $10^{-3}$& $5.736755\times$ $10^{-5}$ \\ \hline

\multirow{9}{*}{$\bf{\alpha_{1} Q + \beta_{1}}$} 
& \multirow{3}{*}{0.9} & -0.001$r_g^{-2}$ & 0.686260 & 0.276581 & -0.000941 & -0.000350 \\
&                      & 0$r_g^{-2}$     & 0.520015 & 0.354605 & -0.010361 & 0.000132 \\
&                      & 0.001$r_g^{-2}$ & 0.529719 & 0.342914 & -0.007294 & 0.000053 \\
\cline{2-7}
& \multirow{3}{*}{1.0} & -0.001$r_g^{-2}$ & 0.793528 & 0.197949 & 0.002911 & -0.000356 \\
&                      & 0$r_g^{-2}$     & 0.602993 & 0.291807 & -0.007044 & 0.000059 \\
&                      & 0.001$r_g^{-2}$ & 0.598763 & 0.288188 & -0.005366 & 0.000030 \\
\cline{2-7}
& \multirow{3}{*}{1.1} & -0.001$r_g^{-2}$ & 0.726749 & 0.196726 & 0.000799 & -0.000252 \\
&                      & 0$r_g^{-2}$     & 0.646310 & 0.248035 & -0.005197 & 0.000028 \\
&                      & 0.001$r_g^{-2}$ & 0.641519 & 0.245924 & -0.003969 & 0.000014 \\
\hline

\multirow{2}{*}{$\bf{Q + \alpha_{2}~Q^2}$}
& $r_g^2$   & - & 0.035807 & 0.450606 & -0.021705 & 0.000496 \\
& $5 r_g^2$ & - & 0.731662 & 0.238435 & -0.002533 & -0.000057 \\
\hline

\multirow{8}{*}{$\bf{Q + \alpha_{3}~e^{\beta_{3} Q}}$}
& \multirow{2}{*}{-0.01$r_g^2$} & 0.1$r_g^{-2}$ & 0.670638 & 0.293679 & -0.000123 & -0.000206 \\
&                               & 0.3$r_g^{-2}$ & 0.642236 & 0.303468 & -0.000924 & -0.000184 \\
\cline{2-7}
& \multirow{2}{*}{-0.001$r_g^2$} & 0.1$r_g^{-2}$ & 0.721506 & 0.246106 & -0.001012 & -0.000159 \\
&                                & 0.3$r_g^{-2}$ & 0.686420 & 0.257592 & -0.002203 & -0.000120 \\
\cline{2-7}
& \multirow{2}{*}{0.001$r_g^2$}  & 0.1$r_g^{-2}$ & 0.728830 & 0.236659 & -0.001292 & -0.000148 \\
&                                & 0.3$r_g^{-2}$ & 0.692212 & 0.248485 & -0.002560 & -0.000105 \\
\cline{2-7}
& \multirow{2}{*}{0.01$r_g^2$}   & 0.1$r_g^{-2}$ & 0.739142 & 0.201066 & -0.003122 & -0.000081 \\
&                                & 0.3$r_g^{-2}$ & 0.717103 & 0.204529 & -0.003471 & -0.000075 \\
\hline

\multirow{4}{*}{$\bf{Q - \alpha_{4} \ln(1-\beta_{4} Q)}$}
& \multirow{2}{*}{-0.3$r_g^2$} & 0.1$r_g^{-2}$ & 0.777180 & 0.240809 & 0.000972 & -0.000219 \\
&                               & 0.3$r_g^{-2}$ & 0.822835 & 0.265600 & 0.003017 & -0.000298 \\
\cline{2-7}
& \multirow{2}{*}{0.3$r_g^2$}  & 0.1$r_g^{-2}$ & 0.739591 & 0.221315 & -0.002198 & -0.000107 \\
&                               & 0.3$r_g^{-2}$ & 0.746217 & 0.185908 & -0.003531 & -0.000157 \\
\hline\hline
\end{tabular}
\end{table*}

We present in Table~\ref{tab:fitting_coefficients} the fitting coefficients obtained for various parameter values. The GR fitting coefficients derived in this work differ by approximately 4\% from those reported in~\cite{Breu_2016_459_646}. This level of deviation is reasonable and comparable to the differences found in~\cite{Staykov2016} and~\cite{Sakstein_2017_95_064013}, considering that our analysis employs a significantly larger set of 53 EoS covering a wide range from soft to stiff models. These results therefore demonstrate the consistency and reliability of our numerical implementation. Notably, the fitting coefficients in the linear $f(Q)$ model reduce to their GR counterparts when $\alpha_1 = 1$ and $\beta_1 = 0$, with only a minor deviation of about 0.6\% at high compactness, which further decreases to approximately 0.2\% at low compactness.

\bsp	
\label{lastpage}
\end{document}